\documentclass[twocolumn,aps,prx,amsmath,longbibliography]{revtex4-1}
\usepackage{graphicx}
\usepackage{subfigure}
\usepackage{hyperref} 
\usepackage{bm}
\usepackage{float}
\usepackage{yfonts}
\usepackage{xcolor}
\hypersetup{
    colorlinks,
    linkcolor={blue!80!black},
    citecolor={blue!80!black},
    urlcolor={blue!80!black}
}
\usepackage[mathcal]{eucal}

\begin{document}

\title{Quench dynamics in superconducting nanojunctions: metastability and dynamical Yang-Lee zeros}
\author{R. Seoane Souto, A. Mart\'{\i}n-Rodero and A. Levy Yeyati}
\affiliation{Departamento de F\'{i}sica Te\'{o}rica de la Materia Condensada,\\
Condensed Matter Physics Center (IFIMAC) and Instituto Nicol\'{a}s Cabrera,
Universidad Aut\'{o}noma de Madrid E-28049 Madrid, Spain}
\date{\today}

\begin{abstract}
We study the charge transfer dynamics following the formation of a 
phase or voltage biased superconducting nano-junction using a full counting statistics analysis. 
We demonstrate that the evolution of the zeros of the generating function 
allows one to identify the population of different many
body states much in the same way as the accumulation of Yang-Lee zeros 
of the partition function in equilibrium statistical mechanics is connected
to phase transitions. We give an exact expression connecting
the dynamical zeros to the charge transfer cumulants and discuss when 
an approximation based on ``dominant" zeros is valid.
We show that, for generic values of the parameters, the
system gets trapped into a metastable state characterized by a non-equilibrium
population of the many body states which is dependent on the initial conditions.
We study in particular the effect of the switching rates in the dynamics showing that,
in contrast to intuition, the deviation from thermal equilibrium increases for
the slower rates. In the voltage biased case the steady state is reached 
independently of the initial conditions. Our method allows us to obtain accurate
results for the steady state current and noise in quantitative agreement with 
steady state methods developed to describe the multiple 
Andreev reflections regime. Finally, we discuss the system dynamics after a sudden 
voltage drop showing the possibility of tuning the many body states population 
by an appropriate choice of the initial voltage, providing a feasible experimental
way to access the quench dynamics and control the state of the system.
\end{abstract}

\maketitle

\section{Introduction}
\label{sec:intro}
The physics of superconducting devices is receiving a renewed attention in parallel with ongoing proposals of applications in quantum technologies 
\cite{Devoret_Science}. While most common designs are based on conventional tunnel junctions, proposals based on hybrid nanostructures like 
those being explored in the search of Majorana bound states are generating a great research activity \cite{Alicea_RPP,Beenakker_RMP}.\\

In these devices a challenging issue is to avoid decoherence for certain low energy states while at the same time being able to 
manipulate them coherently by means of external fields \cite{Flensberg_NJP2017}. An important source of decoherence arises from 
quasiparticle tunneling \cite{Martinis_PRL2009,Catelani_PRB2011,Riste_NatCom2013,Avriller_PRL2015}.
The so-called quasiparticle ``poisoning" can become an obstacle towards the implementation of Majorana qubits \cite{Rainis_PRB,Colbert_PRB2014,Bespalov_PRL2016,Flensberg_PRL2017}. 
Conversely, long lived states arising from trapped quasiparticles in Andreev bound states
(ABS) \cite{Zgirski_PRL,Olivares_PRB2014} have been suggested as possible realizations of a spin 
qubit \cite{Padurario_EPL}.\\

Superconducting nanodevices are also of fundamental interest as an example of an interacting open quantum system, which can be driven out of 
equilibrium by different means and can exhibit highly non-trivial dynamical behavior \cite{Nazarov_book}. While the theory has traditionally focused on the stationary
transport properties, advances in single electron sources and detection techniques are allowing to explore the response of nanodevices in the time domain over increasingly
smaller time scales \cite{FeveScience,BocquillonScience,DuboisNature}.
Moreover, it is becoming clear that the dynamics of open quantum systems can exhibit singular features which are not necessarily reflected
in their stationary properties \cite{Garrahan_PRL2010,Heyl_PRL2013,Karrasch_PRB2013}. These features can be revealed from the full counting statistics (FCS) analysis 
of time-integrated 
observables \cite{Garrahan_PRB2013}. The analogy between equilibrium statistical mechanics and FCS methods suggests that the behavior of the zeros of
the generating function in FCS theory could allow to identify dynamical transitions much in the same way as the Yang-Lee zeros \cite{Lee-Yang1,Lee-Yang2}
of the partition function are connected to phase transitions in the static case \cite{Utsumi_PRB,Peng_PRL,Ivanov_PRE}.\\

In a recent work we have presented a FCS analysis of the quench dynamics in the formation of a superconducting nanojuction \cite{SNS_Souto}. We showed
that, under rather general conditions, many body states with different parity
get a significant population and that their relaxation towards thermal equilibrium 
requires the interaction with external degrees of freedom. In the present work we discuss the phenomenon
from the broader perspective which is provided by analyzing the dynamics of the
FCS Yang-Lee zeros. In contrast to previous works in this direction \cite{Garrahan_PRL2013,Garrahan_PRE2014,Garrahan_PRL2017,Flindt_PNAS}, we 
consider the system evolution at time scales shorter than the typical Markovian times \cite{Zgirski_PRL}.
We study the connection between the structure of the dynamical
Yang-Lee zeros (DYLZ) in the complex plane and the formation of metastable
many body states. Moreover, we show that by tuning a counting parameter, the current cumulants tend to exhibit a singular behavior which is reminiscent of 
the divergences of correlation functions in a first order phase transition.

The paper is organized as follows: In Sec. \ref{sec::model} we introduce the model
used for describing the dynamics of a nanoscale normal region coupled to superconducting leads and
discuss the FCS formalism. We pay particular attention to the definition of the
DYLZ within this context and their connection to the current cumulants.  
In Sec. \ref{sec::initial_conditions} we explore in detail the transient dynamics in quantities like the mean
charge and current. The influence of the initial conditions in the 
formation of ABSs and their effect in the system charge and current evolution is also analyzed, showing 
that it decreases with an increasing coupling to the leads. 
We also explore in this section the effect of the switching rate in the contact formation.
Sec. \ref{sec::FCS_and_DYLZs} is devoted to the FCS analysis of the 
quench dynamics. We show how FCS predicts an evolution from a Poissonian distribution
at short times into a three-modal distribution at larger times which can be
associated to the formation of three different many body states. We discuss 
how a {\it coarse grained} representation can be defined and how the population
of the many body states can be extracted from it. We then give the analysis of the 
evolution of the DYLZ showing how the different many body states can be 
identified from their accumulation in the complex plane. It is also shown that
the scaling of the current cumulants at large times can be extracted from the
dominant DYLZ. The analysis is then extended to  the case of voltage biased junctions 
(Sec. \ref{sec::Vbias}),
discussing how the steady state is reached for quantities such as current and 
noise. This is also illustrated from the evolution of the subgap spectral densities. 
Furthermore, we study the FCS and analyze the particular accumulation of
DYLZ for this voltage biased case. Finally, in Sec. \ref{sec::dc_pulse} we 
analyze the case of a different initialization procedure consisting in a dc voltage 
switch off, demonstrating the possibility of controlling the population of the
different many body states by a proper selection of the applied bias.
Sec. \ref{sec::conclusions} is devoted to some concluding remarks. \\

\section{Model and formalism}
\label{sec::model}
Our model consists of a central region represented by a spin-degenerate quantum level, coupled to two BCS superconducting electrodes. Low energy
electron transport in this kind of structure is dominated by multiple Andreev reflections, leading to the formation of subgap states, located at $\pm\epsilon_A$ 
in the zero bias limit. 
The aim of the present work is the analysis of the transient transport properties through the system after a sudden connection at $t=0$ of the central region to 
the electrodes, which could be phase or voltage biased.\newline 

The system Hamiltonian, $H=H_{leads}+H_{0}+H_T$,
 can be written in terms of Nambu spinors $\hat{\Psi}^{\dagger}_j=\left(c^{\dagger}_{j\uparrow},c_{j\downarrow}\right)$, where 
$j=k\nu,0$ denotes the $\nu=L,R$ lead and the central level states respectively. 
The uncoupled Hamiltonians are given by $H_{0}=\hat{\Psi}^{\dagger}_{0}\hat{h}_{0}\hat{\Psi}_{0}$, 
$H_{leads}=\sum_{k\nu}\hat{\Psi}^{\dagger}_{k\nu}\hat{h}_{k\nu}\hat{\Psi}_{k\nu}$, while the tunneling term is $H_T=\sum_{k,\nu}\left[\hat{\Psi}^{\dagger}_{k\nu}\hat{V}_{\nu}(t)\hat{\Psi}_{0}+\mbox{h.c.}\right]$, where 
$\hat{h}_{0}=\epsilon_0\sigma_z$ and $\hat{h}_{k\nu}= \epsilon_{k\nu}\sigma_z + \Delta_{\nu}\sigma_x$ ($\sigma_z$ and $\sigma_x$ denote here Pauli matrices in the Nambu space). The superconducting gap
parameter will be taken equal for both electrodes, $\Delta_L=\Delta_R\equiv\Delta$, and used as the energy unit. For describing the connection between the system and the electrodes
we use $\hat{V}_{\nu}(t) = f(t) V^0_{\nu}\sigma_z e^{i\sigma_z\phi_{\nu}(t)/2}$, where $f(t)$ is a function controlling the abruptness of the connection as discussed below and 
 $\phi_{L}(t)-\phi_{R}(t) = \phi(t)$ determines the phase difference between the
leads. 

For simplicity we consider a constant normal density of states $\rho_{L,R}$ in the leads with a finite bandwidth $W$ taken as the largest energy scale in the model. We
define the stationary tunneling rates as $\Gamma_{\nu} = \pi (V^0_{\nu})^2 \rho_{\nu}$, and $\Gamma=\Gamma_L+\Gamma_R$. For later use, we also define the normal transmission coefficient 
as $\tau=4\Gamma_L\Gamma_R/(\Gamma^2+\epsilon_{0}^2)$.
 Depending on the relative value between the tunneling rates and the superconducting
gap, two regimes can be identified: the {\it quantum dot} (QD) regime, corresponding to $\Gamma\lesssim\Delta$ and the {\it quantum point contact} (QPC) regime, where
$\Gamma\gg\Delta$. Finally, the central level initial charge will be denoted by $n_{\sigma}(0)$, where $\sigma\equiv\uparrow,\downarrow$. Hereafter we assume 
$\hbar=e=1$.

The time-dependent transport properties of the system are fully characterized by the 
generating function (GF) defined on the Keldysh contour as \cite{levitov}
\begin{equation}
 Z(\chi,t)= \left\langle T_K exp\left[-i\int_C dt' H_{T,\chi}(t')\right]\right\rangle_0,
 \label{GF}
\end{equation}
where $T_K$ is the contour time order operator,
$\chi \equiv \chi_{\nu}(t)$ are counting fields entering as phase factors 
modulating the hopping terms in $H_T$,
having opposite values $\pm\chi_{\nu}$ on the two branches of the Keldysh 
contour. The average in Eq. (\ref{GF}) is taken over the decoupled system. 
The GF gives access to the charge transfer cumulants, i.e. 
$C^n(t) = (i)^n \partial^n \mathcal{S}/\partial \chi^n\rfloor_0$, 
where $\mathcal{S}(\chi,t) = \ln Z(\chi,t)$.
The charge cumulants through the left (right) electrodes can be computed by imposing $\chi_L=\chi$ and $\chi_R=0$ ($\chi_L=0$ and $\chi_R=-\chi$), and for the 
symmetrized charge cumulants $\chi_L=\chi/2$ and $\chi_R=-\chi/2$. The 
corresponding current cumulants are given by $I^n(t) = \partial C^n/\partial t$. The symmetrized cumulants will be denoted $\left\langle I^n\right\rangle$, using 
$\left\langle S\right\rangle=\left\langle I^2\right\rangle$ for the symmetrized shot noise.
As shown in \cite{CohenMillis_PRL2014,CohenMillis_PRB2016}, the occupied Density Of States (DOS) in the transient regime
can be computed from the current to an empty normal electrode, weakly coupled to the central region.
  
It can be shown that $Z(\chi,t)$ can
be computed as a Fredholm determinant on the Keldysh contour \cite{kamenev_book,mukamel,Tang1,Tang2,Souto_PRB2015}. A straightforward extension of this formalism to 
the superconducting case \cite{SNS_Souto} leads to
\begin{equation}
 Z(\chi,t)=\mbox{det}\left[{\bf G}(\chi=0){\bf G}(\chi)^{-1} \right],
 \label{fredholm}
\end{equation}
 where ${\bf G} = -i \left\langle T_K \Psi_{0}(t) \Psi_0^{\dagger}(t') \right\rangle$ is the Green function of the dot coupled to the leads defined in Keldysh-Nambu space.
Using the Dyson equation, Eq. (\ref{fredholm}) can be written as
\begin{equation}
Z(\chi,t) = \det \left[ {\bf G} \left({\bf g}_0^{-1} - \tilde{\bf \Sigma}  \right) \right] \; ,
\label{GF-nambu}
\end{equation}
where ${\bf g}_0$ denotes the uncoupled central level Green function and $\tilde{\bf \Sigma}= \tilde{\bf \Sigma}_L + \tilde{\bf \Sigma}_R$ corresponds to the leads self-energy in which the counting 
field $\chi$ is included. The Keldysh-Nambu components of the self-energy are given by
\begin{eqnarray}
\tilde{\Sigma}^{\alpha\beta}_{\nu,jk}(t,t') = f(t)f(t') s_{\alpha}s_{\beta}s_j s_k \left(V^0_{\nu}\right)^2 e^{i(s_j-s_k)\phi_{\nu}/2}\nonumber\\
 e^{i(s_{\alpha}s_j-s_{\beta}s_k)\chi_{\nu}/2} g^{\alpha\beta}_{jk}(t-t'),
\end{eqnarray}
where $\alpha\beta \equiv \pm$ are the Keldysh indexes, $j\,k \equiv \pm$ are the Nambu ones, $\nu=L,R$ denote the leads,
$s_{\pm}=\pm 1$, $s_{j}= (-1)^{(j+1)}$ and $ g^{\alpha\beta}_{jk}(t-t')$ are the uncoupled leads Green functions. 
Eq. (\ref{fredholm}) has to be integrated numerically by discretizing the Keldsyh contour as depicted in Fig. \ref{keldysh_contour} (for details see
Supplemental Material in Ref. \cite{SNS_Souto}).
Analytical results, which can be obtained in certain limits, will allow us to further clarify our findings as described below.\newline

On the other hand, the GF can be decomposed as 
\begin{equation}
 Z(\chi,t) = \sum_n P_n(t) e^{i\chi n}\;,
 \label{GF_def}
\end{equation}
where $P_n(t)$ can be associated with 
the probability of transferring $n$ charges in the measuring time $t$ \cite{Nazarov_book}.
In the superconducting case, the charge in the leads is not well defined, and $P_n(t)$ can eventually 
take negative values \cite{belzig,ramer,clerk}. The $P_n(t)$ are therefore referred to in this case as {\it quasi-probabilities}.


\begin{figure}
\includegraphics[width=.8\linewidth]{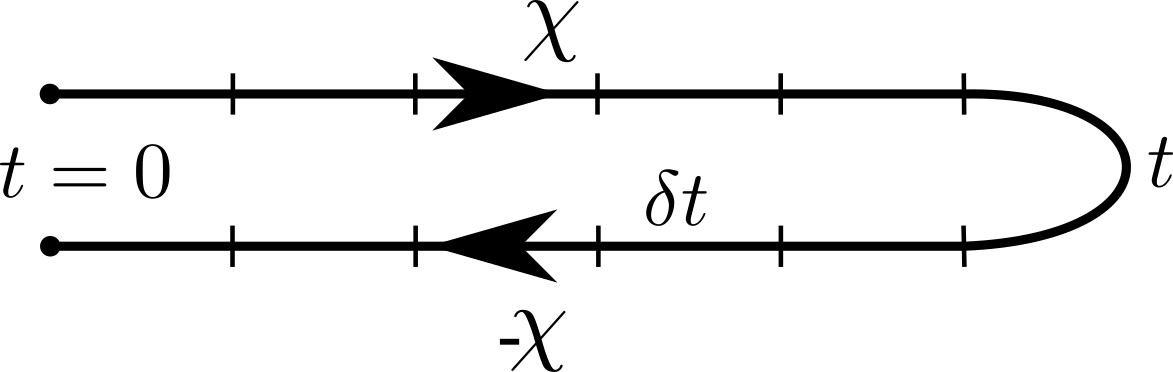}
\caption{Keldysh contour considered to analyze 
the transient regime. $\chi$ indicates the counting field changing sign on the two branches of the contour and $\delta t$ corresponds to the time step in the discretized calculation of the dot Green 
function and the generating function $Z(\chi,t)$.}
\label{keldysh_contour}
\end{figure}



\subsection{Relation between DYLZs and cumulants}
In their seminal papers, T. D. Lee and C. N. Yang demonstrated the connection between thermodynamical phase transitions and the behavior of
the roots of the partition function \cite{Lee-Yang1,Lee-Yang2}. They discussed how these roots accumulate to form branches in the complex 
plane of a given variable, $z$, dependent on the system's 
temperature. 
The phase diagram of the system is determined by the interceptions of these branches with the positive real axis in the thermodynamical limit 
(i.e. as the volume tends to infinity). The crossing points 
correspond to situations where two (or more phases) coexist. These ideas, originally developed for equilibrium statistical mechanics, have recently been applied to the study of 
the time evolution of open quantum systems \cite{Garrahan_Flindt,Garrahan_PRL2017}, with the time playing the extensive role of the volume, the GF of Eq. (\ref{GF_def}) the role of the partition 
function and $z \equiv e^{i\chi}$.
By analogy, the roots of the GF in the complex $z$ plane are referred to as Dynamical Yang-Lee Zeros (DYLZs). The position of the DYLZs, denoted by $z_j(t)$, 
fully characterize the transport properties through the system, and Eq. (\ref{GF_def}) can be rewritten as
\begin{equation}
 Z(z,t)=\prod_{j}(z-z_j(t)) \;.
\end{equation}

In a previous work we derived exact expressions for the charge cumulants of arbitrary order \cite{Souto_Fortschritte}. In terms of the DYLZs these can be written as

\begin{equation}
C^n(t)=-\sum_{j}\mbox{Li}_{1-n}\left(\frac{1}{z_j(t)}\right) 
 \label{polylog}
\end{equation}
where Li$_j$ denotes the polylogarithm function of order $j$ \cite{abramowitz_stegun}. The main contribution to the charge cumulants is provided by the DYLZs close 
to $z=1$, where the functions diverge. For the higher order cumulants, the exact expression Eq. (\ref{polylog}) can be well approximated by 
\cite{Berry,Flindt_PRB2010,Kambly_factorial}
\begin{equation}
C^n(t)\approx (-1)^{n-1}(n-1)!\sum_j\frac{2\cos\left\{n\,\mbox{arg}\left[z_j(t)-1\right]\right\}}{\left|z_j(t)-1\right|^n}\,.
\end{equation}

Additional information can be obtained from the so-called {\it factorial cumulants}, which are a generalization of the conventional ones, defined by shifting the measurement 
point in the z-plane. These quantities
provide valuable information about the interactions in mesoscopic systems \cite{Kambly_factorial,Stegmann_factorial1,Stegmann_factorial2}. The factorial generating function (FGF) can 
be written as 
\begin{equation}
 Z_F(z,t;s)=\frac{\sum_n P_n \left(z+s\right)^n}{\sum_n s^n P_n}\;,
\end{equation}
where $s$ is a biasing field. Notice that the denominator is a normalization factor and does not contribute to the transport properties, 
since it does not depend on the counting field. One can also define the DYLZs of the FGF, which are just the original $z_j(t)$ shifted by $s$. Thus, the factorial cumulants are given by
\begin{equation}
  C^{n}_F(t;s)=-\sum_{j} \mbox{Li}_{1-n}\left[\frac{1}{z_j(t)+s}\right]\;,
 \label{polylog_fact}
\end{equation}

The current cumulants ($I^n$) and the factorial current cumulants ($I^{n}_F$) can be then computed by deriving Eqs. (\ref{polylog}) and (\ref{polylog_fact}) with respect to time, respectively.
Again we shall define $\left\langle I^{2}_F\right\rangle=\left\langle S_F\right\rangle$ for the symmetrized factorial shot noise.

\section{Transient dynamics}
\label{sec::initial_conditions}

In this section we analyze the time evolution of single particle observables (charge, current and spectral densities) after a sudden switch on of the central level-leads
coupling for the phase biased case. Parameters are chosen in order to study the different behavior from the QD to the QPC regimes. Unless stated differently, we 
consider in this section the electron hole-symmetric case $\epsilon_0 =0$ and $\Gamma_L=\Gamma_R$. This choice corresponds to a case of perfect transmission where the nonequilibrium effects that
we are interested in are more pronounced.

\subsection{Central level charge evolution and ABSs formation}
\label{subsec::charge}

\begin{figure}
\includegraphics[width=1\linewidth]{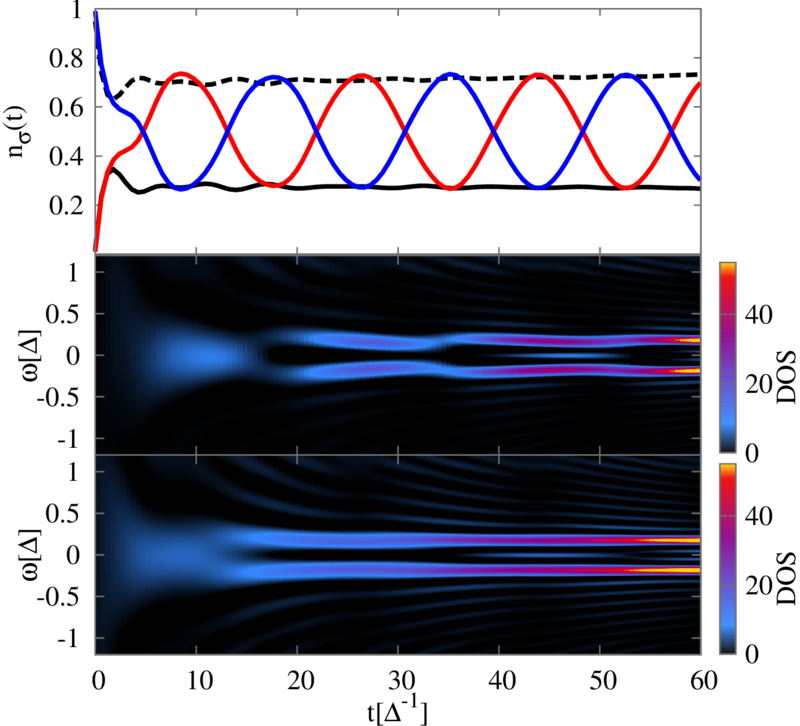}
\caption{Top panel: Time evolution of the central level population for three different initial configurations:  $(n_\uparrow(0),n_\downarrow(0))=(0,0)$ (red line), $(1,1)$ (blue line), where 
$n_\uparrow(t)=n_\downarrow(t)$, and $(0,1)$ 
(black line, using solid line for $n_\uparrow(t)$, and the dashed one for $n_\downarrow(t)$). Lower panels: time evolution of the occupied density of states (DOS) for the (0,0) and (0,1)
initial configurations. We consider a perfect transmitting junction in the quantum dot regime with $\Gamma=0.5$ (in units of $\Delta$) and $\phi=2$.}
\label{charge_evolution_spectral}
\end{figure}

\begin{figure}
\includegraphics[width=1\linewidth]{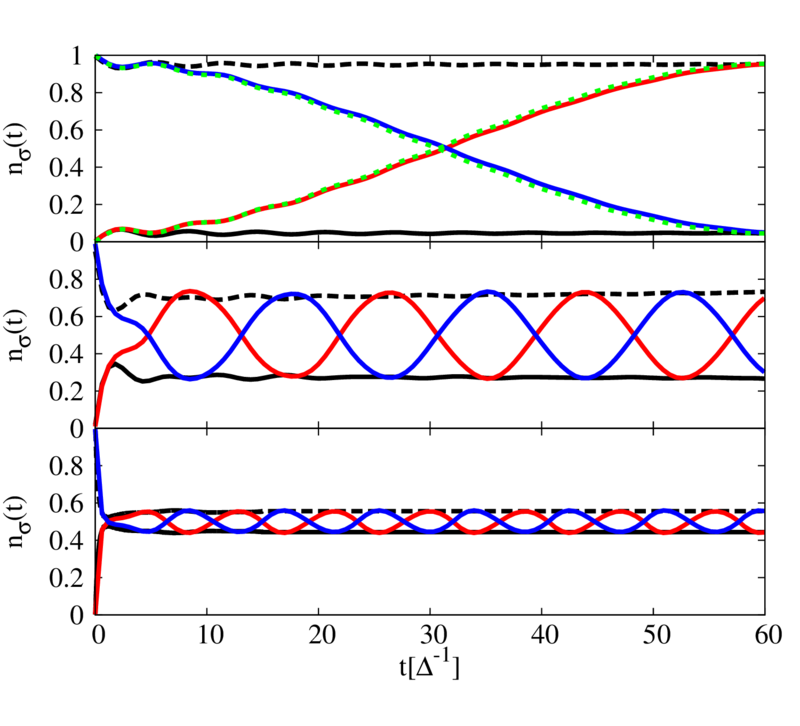}
\caption{Time evolution of the level population for three different initial configurations:  $(n_\uparrow(0),n_\downarrow(0))=(0,0)$ (red line), $(1,1)$ (blue line) and $(0,1)$ 
(black line, using solid line for the population of spin up electrons, and the dashed one for spin down). The dotted line in the upper panel shows the comparison to the approximation
described in the appendix \ref{sec::SPA}. We show the evolution of the behavior of the population for three different couplings between the central level and the electrodes $\Gamma=0.05$, $0.5$ and 
$2$, from top to bottom, for a phase difference $\phi=2$.}
\label{charge_evolution}
\end{figure}

In Fig. \ref{charge_evolution_spectral} we show results for the charge per spin and the subgap occupied spectral density  
evolution after a sudden connection, i.e. $f(t)=\theta(t)$, for three 
different initial configurations, $(n_\uparrow(0),n_\downarrow(0))=(0,0), (1,1)$ and $(0,1)$  and $\Gamma=\Delta/2$. At short times, $t\lesssim1/\Delta$, the initial 
excess charge tends to relax through the 
electrodes. A change in this tendency is observed at times of the order $t\sim2/\Delta$ coinciding with the incipient formation of the ABSs inside the gap
which block the excess charge relaxation.  While for the initial 
configurations $(0,1)$ or $(1,0)$ the system gets trapped in a metastable magnetic state, with $n_{\uparrow}
\ne n_{\downarrow}$, for the initial configurations $(0,0)$ and $(1,1)$ the charge oscillates but the system remains non-magnetic. The period of the oscillation
is $\sim\pi/|\epsilon_A|$, where $\epsilon_A \simeq \Gamma \cos(\phi/2)$ corresponds to the ABS energy in the QD regime. It should be also noticed that the oscillations corresponding
to the $(0,0)$ and the $(1,1)$ configurations are displaced in half a period. Remarkably, as we show in the next subsection, in all these cases the system exhibits the same symmetrized current.\newline
 
As shown in Ref. \cite{SNS_Souto}, one can get an analytical insight on this behavior as the spectral weight in this QD regime is mainly concentrated on the ABSs
and the retarded dot Green function can be approximated by just the contribution from these states. In Ref. \cite{SNS_Souto} we analyzed the charge evolution starting
from the initial magnetic configuration $(0,1)$. The corresponding analysis for an arbitrary initial configuration is given in Appendix \ref{sec::SPA}. As we show
in this Appendix the dot charge oscillations for the $(0,0)$ case can be approximated as $2|P_{12}|^2[1-\cos(2\epsilon_A t)]$, where
\begin{equation}
 |P_{12}|^2=\frac{\left(1-\Gamma/\Delta\right)^2-\epsilon_{0}^2/\epsilon_{A}^2}{4\left(1+\epsilon_{A}^2/\Delta^2\right)}\,.
\end{equation}
These oscillations remain undamped unless an additional relaxation mechanism is
included. The amplitude is determined by the coupling between the two ABSs ($P_{12}$), generated by the initial conditions. This behavior is also reflected in the occupied DOS, shown in the middle 
panel of Fig. \ref{charge_evolution_spectral} for the $(0,0)$ initial condition. We show that the dot's charge oscillations are correlated with an intermittent behavior of the population of the ABSs. 
This behavior is absent in the half-filled initial condition (lower panel in Fig. \ref{charge_evolution_spectral}).\newline


In Fig. \ref{charge_evolution} we show the charge evolution for the same initial configurations studied in Fig. \ref{charge_evolution_spectral} for three different couplings to the electrodes, 
$\Gamma=0.05$, $0.5$ and $2$ (in units of $\Delta$), from top to bottom. As commented before in the QD regime, $\Gamma\lesssim\Delta$, for initially trapped quasi-particles
(i.e. $(0,0)$ and $(1,1)$ configurations), the population exhibits large oscillations. The amplitude of these oscillations is monotonously reduced when increasing the hybridization, $\Gamma$. In the QPC
regime, $\Gamma\gg\Delta$, the initial condition is almost fully relaxed at very short times ($t \sim 1/\Gamma)$ and the population tends to reach the
expected stationary value $n_{\sigma} \sim 0.5$. However, as shown in Ref \cite{SNS_Souto}, this relaxation of the initial excess charge does not imply
a full thermalization of the system. This would be further analyzed in Sec. \ref{sec::FCS_and_DYLZs}.

\subsection{Current evolution}
\begin{figure}
\includegraphics[width=1\linewidth]{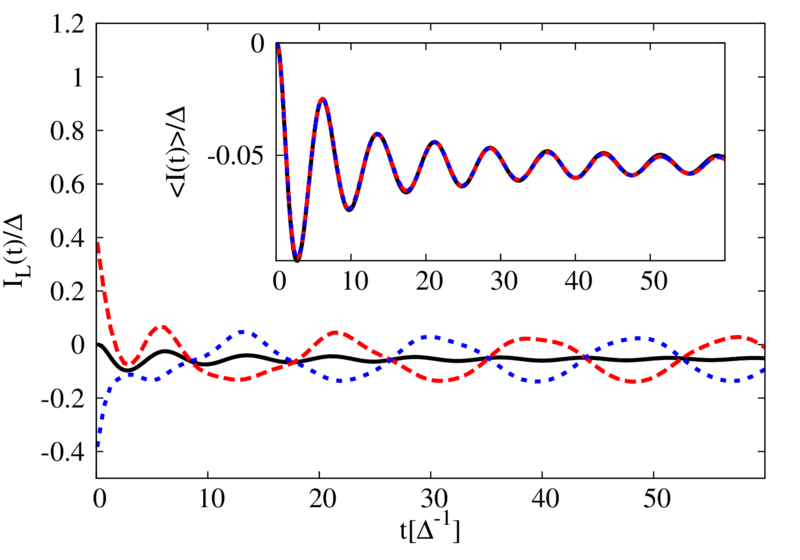}
\caption{Left current evolution for three different initial configurations $(n_\uparrow(0),n_\downarrow(0))=(0,0)$ (dashed red line), $(1,1)$ (dotted blue line) and $(0,1)$ (solid black line). The parameters
are the same as in Fig. \ref{charge_evolution_spectral}. In the inset we show the independence on the initial conditions of the symmetrized current.}
\label{current_sym}
\end{figure}

In this subsection we analyze the main results for the transient current flowing through the system after a sudden contact formation. In the main panel of Fig. \ref{current_sym} we show the current 
evolution at the left interface for the three initial configurations studied before. For the case with an initially trapped quasiparticle (blue and red curves), this current exhibits similar oscillations as found in
the charge evolution, while for the $(0,1)$ case it approaches the mean value of the previous two cases (solid black like). 
On the other hand, the transient current becomes independent on the initial
charge configuration when it is left-right symmetrized, as shown in the inset of Fig. \ref{current_sym}. This fact demonstrates that most of the oscillatory behavior 
arises from the symmetric transfer of quasiparticles between the central region 
and the electrodes, which cancel out when the current is symmetrized.
From now on we will focus on the properties of the symmetrized current.\newline

\begin{figure}
\includegraphics[width=1\linewidth]{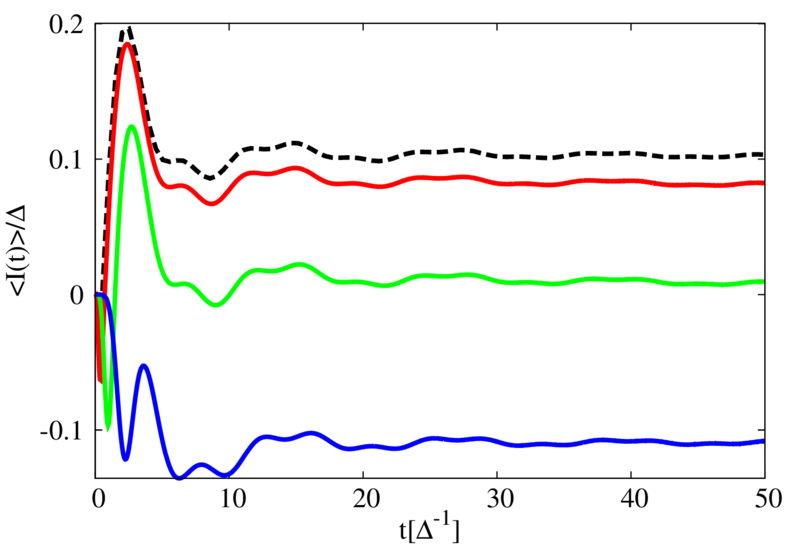}
\caption{Transient current as a function of the switching rate $\alpha$ of the coupling between the electrodes and the central region (see text) for the case  
$\Gamma_L=\Gamma_R=5$. The dashed black curve shows the quench evolution, compared with three different connection rates,
$\alpha=5$, $1$ and $0.25$, from top to bottom. The rest of parameters are the same as Fig.
\ref{charge_evolution_spectral}.}
\label{smoothSwitch}
\end{figure}

Another characteristic of the current flowing through the system is that the long-time asymptotic value does not reach the expected limit for a thermal equilibrium situation. The characterization 
of this metastable state has already been done in Ref. \cite{SNS_Souto} for a sudden quench of the coupling to the electrodes. An issue not addressed in that work was the effect of a decreasing
switching rate. In Fig. \ref{smoothSwitch} we show the current evolution in the point contact limit ($\Gamma\gg\Delta$) assuming $f(t)=\theta(t)[1-\exp(-\alpha t)]$, $\alpha$ being the 
connection rate.
For a sufficiently fast connection, we recover the results of Ref.
\cite{SNS_Souto} (dashed black curve). Surprisingly, for slower connection rates, the system gets trapped in a metastable state which deviates more strongly from the equilibrium situation, with a supercurrent 
which is even inverted for the smaller connection rates. This indicates that the trapping of the system in a 
metastable state is not an artifact of 
the abrupt connection, but is a rather general result. The behavior is better understood from the discussion of the Andreev states population in the next section.

\subsection{Andreev states population}
\begin{figure}
\includegraphics[width=1\linewidth]{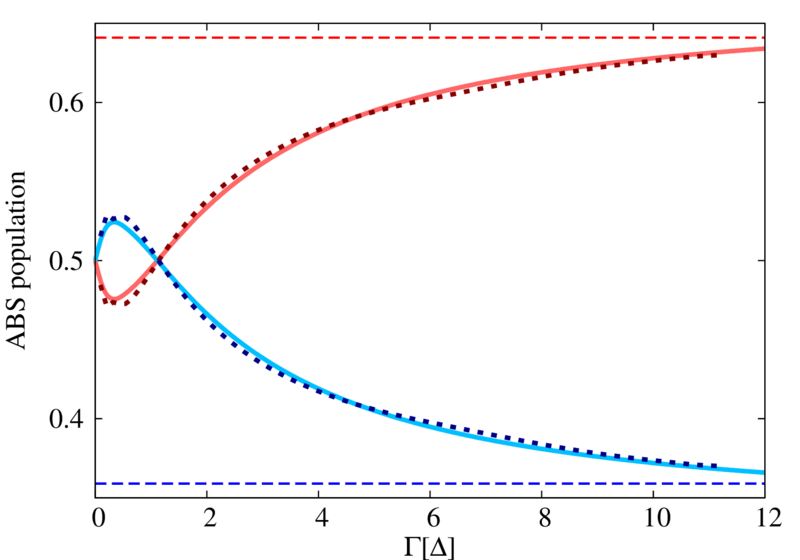}
\caption{Occupation of the lower (red) and upper (blue) ABSs for the perfect transmitting case ($\epsilon=0$ and $\Gamma_L=\Gamma_R=\Gamma/2$) and $\phi=2$ as a function of $\Gamma$. The dashed 
lines show the asymptotic value in the point contact regime, given in Appendix \ref{sec::rate_eq}. The dotted line represents the long time population varying the connection rate $\alpha$, as a function
of the effective tunneling rate $\Gamma_{eff}=\Gamma(1-(1-exp(-\alpha t_A))/\alpha t_A)$, with $t_A=\sqrt{2}/\Delta$ and $\Gamma=12$.}
\label{AndreevOcupation}
\end{figure}


A property that can be accessed from the transient current behavior is the population of the two ABSs. Using the symmetrized current and the fact that, in average, the two ABSs have to be half-filled, 
their population can be extracted. The set of equations used to extract the population are
\begin{eqnarray}
 \left\langle I\right\rangle&=&(n_{d}-n_{u})I_A+I_c\nonumber\\
 1&=&n_{d}+n_{u} \;,
\label{ABS_pop_analytics}
\end{eqnarray}
$n_{u}$ and $n_d$ being the population of the upper and lower ABS respectively, $I_A$ the equilibrium supercurrent supported by the lower ABS and $I_c$ the contribution from the continuum 
to the current. The long time average occupation of the two ABSs after a quench of the 
central region-leads coupling is shown in Fig. \ref{AndreevOcupation}. 
For $\Gamma\ll\Delta$, the upper ABS is more populated than the lower one, leading 
to a current flowing in the opposite direction than the 
expected stationary value. This behavior is inverted at $\Gamma\sim\Delta$, as predicted by the analytical insight described in Appendix \ref{sec::SPA}.
For $\Gamma\gg\Delta$ the system reaches a universal behavior, represented by the discontinuous lines. This universal behavior can be 
extracted from simple rate equations, assuming transition rates of the order of the distance between the continuum and the final states (see Appendix \ref{sec::rate_eq}).\newline

Finally, the dotted lines in Fig. \ref{AndreevOcupation} show the Andreev states population for the smoother connection case, with an effective tunneling rate, $\Gamma_{eff}$, dependent on the connection 
rate $\alpha$. We observe similar features to the sudden quench situation, indicating that the quasiparticle relaxation happens at very short times (before the ABSs formation). After that initial stage, 
the two ABSs move adiabatically to their long time stationary value without exchanging charge.

\section{Full Counting Statistics and Dynamical Yang-Lee Zeros}
\label{sec::FCS_and_DYLZs}

\subsection{Full counting statistics}
\begin{figure}
\includegraphics[width=1\linewidth]{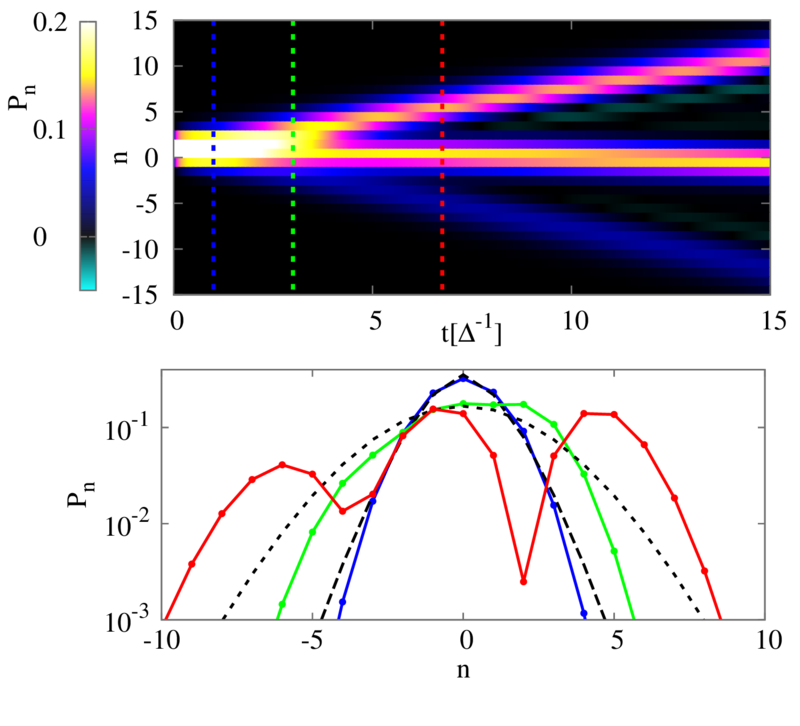}
\caption{Upper panel: Time evolution of the quasiprobabilities, exhibiting the formation of three different states. In the lower panel we show some cuts at different times $t=1$ (blue) , 
$3$ (green) and $6$ (red). At short times, the probabilities can be interpreted as a birth-death distribution (discontinuous line) with equal birth and death rates (see appendix \ref{sec::B-d} for details).
At times of the order of the creation of the ABSs, the distribution tends to deviate from the simple bidirectional poissonian distribution (comparing green curve to dotted one). At longer times 
(red curve), three states are created, related to the many body representation of the 
population of the ABSs.}
\label{short_time_FCS}
\end{figure}

In the QPC regime and for steady state conditions, the quantum state of the system can be characterized through the many body spectrum representation, corresponding 
to the four possible occupations of the ABSs \cite{Zgirski_PRL,Zazunov_PRB2014}. In the ground state ($-$)
only the lower ABS is occupied. An excited state of the same parity  corresponds to populate only the 
upper ABS ($+$). Finally, there are two degenerate excitations involving a change in the parity of the system state, which will be referred to 
as {\it odd} states ($odd$), corresponding to 
populate or depopulate both ABSs simultaneously. While this simplified description does not hold at short times ($t<1/\epsilon_A$) in the 
transient regime, as it was shown in Ref. \cite{SNS_Souto}, the population of these states can be inferred by analyzing 
the evolution of the asymptotic quasiprobabilities. \newline

The upper panel of  Fig. \ref{short_time_FCS} shows the time evolution of the quasiprobabilities at short times, which evolve from a uni-modal distribution to a tri-modal one, related to the three
states described above. In the lower panel of Fig. \ref{short_time_FCS} we show some cuts before (blue) and just after this transition (red). 
At very short times (blue curve)
the charge transfer is a random process that involves charge flowing in both directions, with similar probabilities. 
This short time dynamics can be described as a bidirectional Poisson distribution (see appendix \ref{sec::B-d}), shown as discontinuous lines in the lower panel of Fig. \ref{short_time_FCS}. 
The green line shows the probability distribution at the typical formation time of the ABSs. At this time, the probability distribution becomes 
asymmetric exhibiting a net charge flowing through the junction which deviates from the fitted bidirectional Poisson distribution (dotted curve). This fit provides an estimated number of $\sim3$ 
electrons crossing the junction in each direction to create the subgap states.
At longer times, the distribution exhibits three maxima, indicating the coexistence between the different many body states, which in the following will
be referred to as (quantum) phases, in analogy with equilibrium statistical mechanics.\newline 

\begin{figure}
\includegraphics[width=1\linewidth]{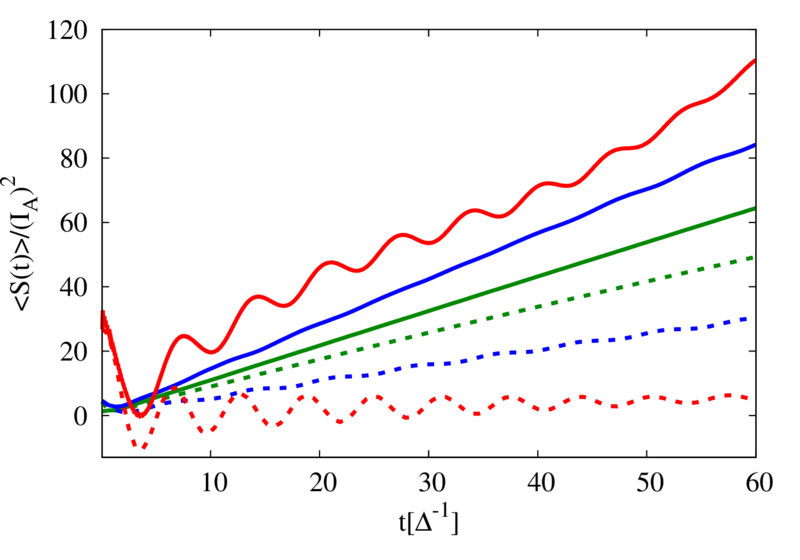}
\caption{Symmetrized shot noise of the transferred charges for different initial conditions $(n_\uparrow(0),n_\downarrow(0))=(0,0)$ or $(1,1)$ (continuous lines) and $(1,0)$ (dashed lines)  
for three different dot electrodes coupling, $\Gamma=0.05$ (red), $0.5$ (blue) and $2$ (green)
showing the sensitivity of the symmetrized noise to the initial conditions}
\label{noise_sym}
\end{figure}

Further insight on the system short time dynamics can be obtained by analyzing the current noise. In Fig. \ref{noise_sym} we show some results for the symmetrized current noise for different couplings 
to the electrodes and initial configurations. 
As can be observed, differently from the symmetrized transient current or the ABS population, Eq. (\ref{ABS_pop_analytics}), the symmetrized noise {\it is} sensitive to 
the initial conditions. 
This fact shows that the actual many body state cannot be inferred solely from mean single particle properties, but requires the knowledge of higher order current cumulants.
The dashed lines correspond to the evolution for an initial condition $(0,1)$, while the solid ones correspond to
the evolution for initially trapped quasiparticles (cases $(0,0)$ and $(1,1)$). 
In all situations, we observe a linear increase of the noise with time, which can be considered as
a signature of the phase coexistence.
The noise becomes larger for the case of initially trapped quasi-particles, 
coinciding with the oscillations observed in the dot population. 
The dependence on the initial conditions 
decreases for increasing $\Gamma$. This dependence on the initial conditions is present also in the many body population of the ABSs, which 
fully characterize the state of the system.

\subsection{Coarse grained statistics}
At long times a simplified coarse grained representation of the FCS can be introduced, where we approximate the probability map as three maxima disregarding their width. Their
weights ($P_-$, $P_+$ and $P_{odd}$) can be computed by integrating the quasiprobabilities around their maxima. The three peaks evolve with time as  
$I_\mu t$ ($\mu=-,+,odd$) with
\begin{equation}
 I_-=-I_+=I_A+I_c \quad \mbox{and}\quad I_{odd}=I_c\;.
\end{equation}
This representation provides an accurate description of the transport properties in the long time regime, where the width of the three probability peaks becomes negligible compared to 
their separation. The long time GF can then be written as
\begin{equation}
 Z(\chi)\approx P_- e^{i\chi I_-\,t}+P_+ e^{i\chi I_+\,t}+P_{odd} e^{i\chi I_{odd}\,t}.
\label{coarsed_GF}
\end{equation}
From this expression, the asymptotic position of the DYLZs can be obtained as
\begin{equation}
 \alpha_{\pm}=z^{I_A t}_\pm\approx\frac{-P_{odd}\pm\sqrt{P_{odd}-4P_-P_+}}{2P_-}\,,
\label{dominant_poles}
\end{equation}
which corresponds to two branches, converging to the unitary circle centered in the coordinate's origin. In the thermodynamical limit, a similar shape has already been reported in Ref. \cite{Lee-Yang2} 
for the Ising model, which describes the system undergoing a phase transition at $z=1$. The point $z=0$ is also a root of the GF with a degeneracy of $I_+t$. Using Eq. (\ref{polylog}), simple
expressions can be derived for the current cumulants, i.e.
\begin{equation}
 \left\langle I^n\right\rangle\approx n\,I_{A}^n\, t^{n-1}\sum_\pm\mbox{Li}_{1-n}\left(\frac{1}{\alpha_\pm}\right)\,,
\end{equation}
which describe the way the cumulants diverge with time. For instance, this equation characterizes the linear increase in the noise shown in Fig. \ref{noise_sym}. As it was pointed out in Ref.
 \cite{SNS_Souto}, the long time current and noise (together with the normalization condition) provide a complete set of equations for determining the population 
of each of the three phases. Extrapolating this reasoning to the case of $p$ coexistent phases, the population of each of the phases could be determined by measuring the long time behavior of the first 
$p-1$ cumulants.

\subsection{Dynamical Yang-Lee zeros}

\begin{figure}
\centering     
\includegraphics[width=1\linewidth]{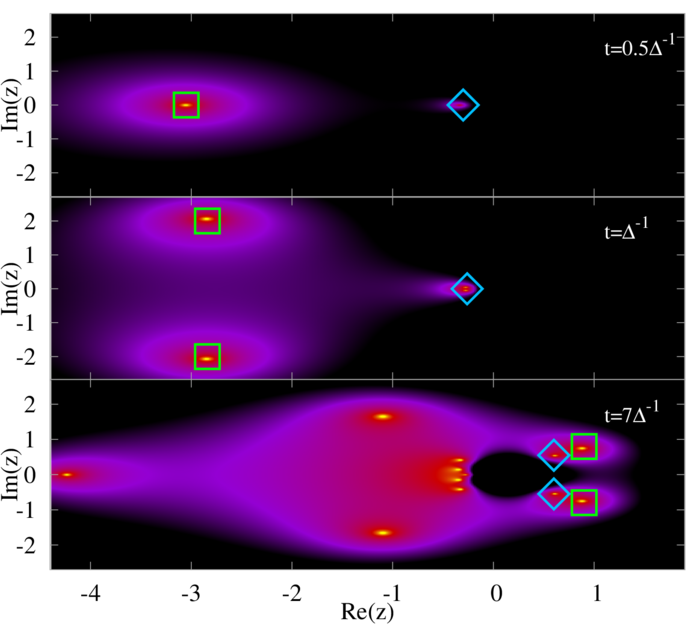}
\caption{Short time behavior of the modulus of the inverse of the GF ($1/|Z(z,t)|$) as a function of the $z$ complex variable. At short times (top panel) we observe the zeros in the real 
negative axis, consistent with a non-interacting situation. At times of the order of $t\sim\Delta^{-1}$, the superconducting correlations become important and the zeros can appear as complex 
conjugate pairs. At longer times, the dominant DYLZs approach the measurement point, $z=1$. For clarity, we use the squares and diamonds to mark the dominant zeros. The parameter values are $\Gamma=2$, 
$\epsilon=0$, $\Delta=1$, $\phi=2$ and times $t=0.5$, $1$ and $7/\Delta$ (from top to bottom).} 
\label{poles:G2}
\end{figure}

\begin{figure}
\includegraphics[width=1\linewidth]{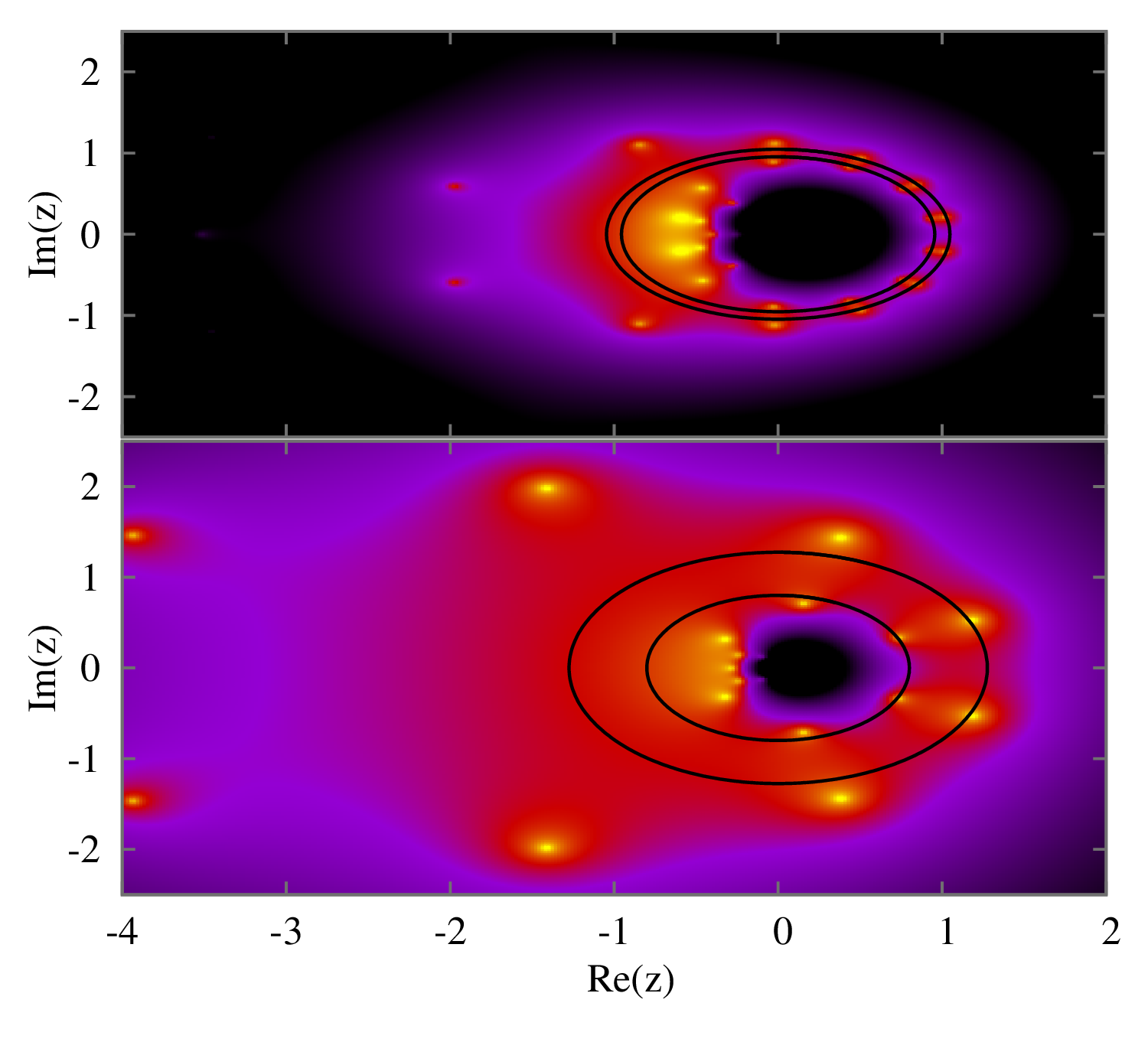}
\caption{Long time $(t\sim25/\Delta)$ behavior of $1/|Z|$ illustrating the accumulation of DYLZs around two circles which can be described by the coarse grained expression of Eq.
(\ref{dominant_poles}). We show two different situations for $\Gamma=2$ (top panel) and $0.5$ (bottom panel). The other parameters are the same as in Fig. \ref{poles:G2}.}
\label{poles_G}
\end{figure}

An alternative approach to the problem is provided by the analysis of the
behavior of the DYLZs, which according to Eq. (\ref{polylog}) fully characterize the transport properties of the system. 
In Fig. \ref{poles:G2} we plot  $1/|Z(z,t)|$, where the bright spots correspond to the DYLZs. At short times ($t\ll1/\Delta$), the zeros are distributed along the negative real axis, a signature of 
uncorrelated electron transport \cite{Abanov_PRL2008,Ivanov_europysics,Kambly_factorial,Utsumi_PRB}. At intermediates times ($t\sim 1/\Delta$), superconducting correlations become important, 
and the zeros appear as complex conjugate pairs, shown by the green squares in the middle panel. At longer times ($t\gg1/\Delta$), two pairs of complex conjugated zeros (represented by the symbols in the 
lower panel of Fig. \ref{poles:G2}) approach the measurement point $z=1$. 
These DYLZs will be referred to as {\it dominant zeros}, since they provide the main contribution to the cumulants given by Eq. (\ref{polylog}).
In Fig. \ref{poles_G} we show the accumulation of the DYLZs in the complex $z$-plane in the long time limit forming branches, for two different values of the tunneling rate. The dominant zeros tend to
accumulate along the coarse grained result, given by Eq. (\ref{dominant_poles}), which describes two concentric circles 
(black lines). The description of the zeros located farther from the origin is poorer, since they may depend on details, such as the peak's width, not included in the simplified model.
The regions delimited by the 
two circles can be associated to states where the system is in a single phase, while the circles describe the phase coexistence lines \cite{Lee-Yang1,Lee-Yang2}. The nature of each of the phases can be 
inferred from their transport properties. In the limit $t\to\infty$, the two circles
tend to converge to the unitary one ($|z|=1$), leading to the coexistence of three phases at the measurement point, $z=1$, which thus becomes a {\it triple point}. This image is consistent with the one
 provided by the quasi-probabilities in Fig. \ref{short_time_FCS}.  We would like to emphasize that the radius of the two circles in Fig. \ref{poles_G} is controlled by the divergences at the
superconducting gap of the leads BCS density of states. A small broadening of these divergences, which, as discussed in Ref. \cite{SNS_Souto},
causes the relaxation of the system towards the 
equilibrium stationary state, reduces the radius of the circle moving the transition point towards smaller $|z|$ values.\\

\begin{figure}
\centering     
\includegraphics[width=1\linewidth]{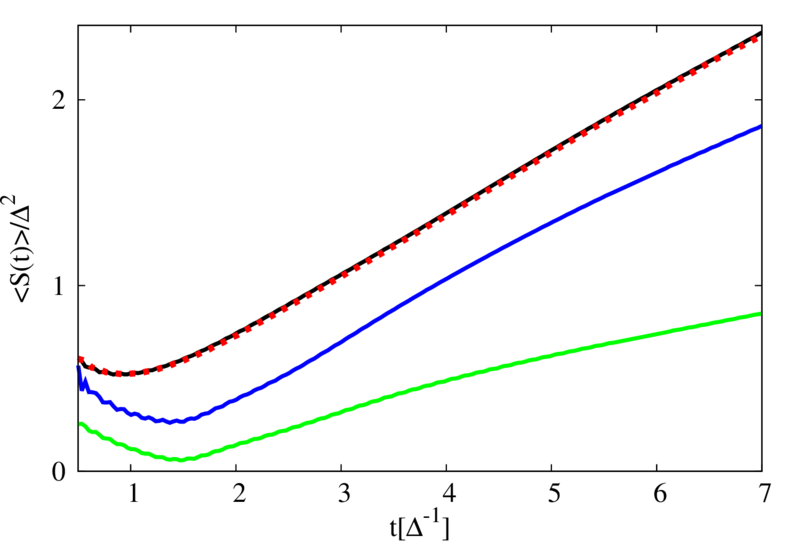}
\caption{Rebuilt noise evolution from the position of the dominant zeros. The green line corresponds to the noise evolution considering only the two dominant zeros (red squares in Fig. \ref{poles:G2}). 
For the blue line we consider 4 zeros (the two dominant and 2 subdominant ones, blue diamonds in Fig. \ref{poles:G2}), recovering the slope of the noise at short times. The dashed red line shows the 
convergence to the full result (black line) when considering a higher number of zeros (in this case around $10$).}
\label{noise_poles_G2}
\end{figure}

In Fig. \ref{noise_poles_G2} we show the shot noise computed from the dominant zeros, using Eq. (\ref{polylog}). The time scaling of the shot noise is well described by the four dominant zeros 
marked with symbols in Fig. \ref{poles:G2}. This result is at
variance with the case analyzed in Ref. \cite{Lee-Yang2}, where only two phases coexist and thus only two dominant zeros are needed. In our case, however, two branches of DYLZs are needed, 
since $z=1$ becomes a triple point when $t\to\infty$.\\

\begin{figure}
\includegraphics[width=1\linewidth]{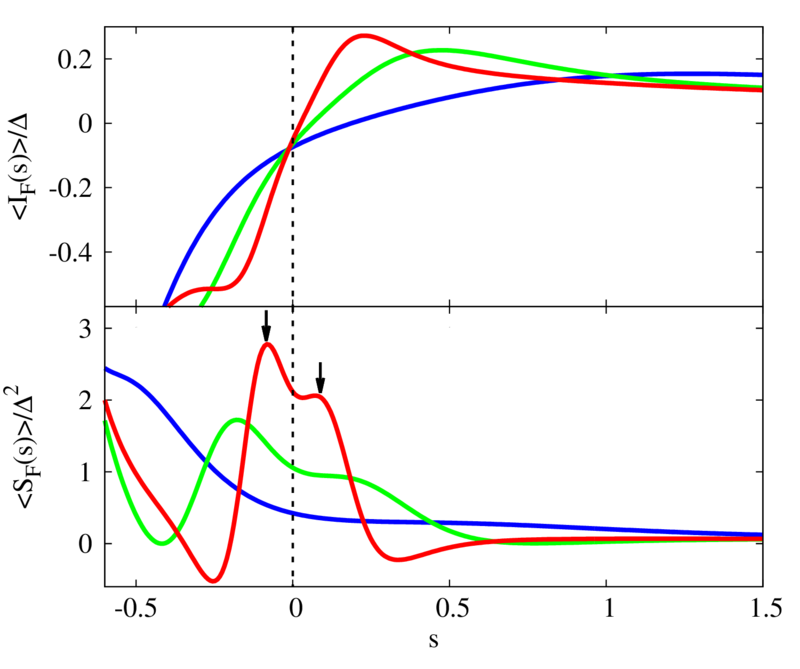}
\caption{Factorial current (top) and noise (bottom panel) as a function of the bias field $s$. We show the evolution for three times, $t=10$ (blue), $25$ (green) and $50$ (red). The factorial current
tends to exhibit jump around the measurement point (black dashed line) while the factorial noise tends to diverge at this point, which can be considered a signature of a dynamical first order transition.
The factorial noise also exhibits two maxima, (marked with the arrows), associated to the intersection between the real $z$ axis and the 
two circles (see Fig. \ref{poles_G}). The parameters are the same as in the lower panel of Fig. \ref{poles_G}.}
\label{factorialCumulants}
\end{figure}

Finally, in Fig. \ref{factorialCumulants} we show the first two factorial cumulants 
(current and noise) as 
a function of the measurement point over the positive real $z$-axis, parametrized by the bias field $s$, see Eq. (\ref{polylog_fact}), for increasing times. 
The parameters are the same as in the lower panel of Fig. \ref{poles_G}. In the factorial current, we observe a tendency to the formation of a jump at the measurement point, $s=0$, indicated by a dashed 
line in Fig. \ref{factorialCumulants}. This figure provides information about the nature of each of the phases: for $s>0$ (outside the two circles in Fig. \ref{poles_G}) the current is positive, which
for the choice of parameters, is a signature of the dominance of the ground state. In contrast, for $s<0$ (inside the two circles) the current is negative, indicating
the dominance of the even excited state, while for $s\simeq 0$, i.e. between the two circles in Fig. \ref{poles_G} the current almost vanishes. 
On the other hand, the factorial noise tends to exhibit two maxima approaching the measurement point ($s\sim0$) marked with arrows in the lower panel of Fig. \ref{factorialCumulants}, corresponding
roughly to the condition $s+\alpha_\pm=1$ associated to the intersections with the phase coexistence lines (i.e. the circles indicated in Fig. \ref{poles_G}). Again the increasing noise for 
$t\to\infty$ is a signature of phase coexistence as shown in Fig. \ref{noise_poles_G2}. Although not shown, there is another divergence at $s\approx-1$ which corresponds to the point $z=0$, 
where a divergence naturally occurs due to the presence of charge transfer processes in the opposite direction to the mean current, see Eq. (\ref{coarsed_GF}).\\

\section{Voltage biased junction}
\label{sec::Vbias}

In this section we summarize the main results when a voltage bias is symmetrically applied to the junction $(\mu_L=-\mu_R=V/2)$.
Some previous works have analyzed time resolved transport in superconducting nanojunctions, although focusing on the single particle properties
\cite{Stefanucci_PRB_2009,Stefanucci_PRB_2010,Albrecht_2013,Weston_PRB_2016,Taranko_arXiv_2017}. The voltage bias can be incorporated in the superconducting phase, using a gauge transformation, leading to 
$\phi_\nu(t)=\phi_\nu(0)+\mu_\nu t$ ($\nu=L,R$).

\subsection{Current evolution}
\begin{figure}
\includegraphics[width=1\linewidth]{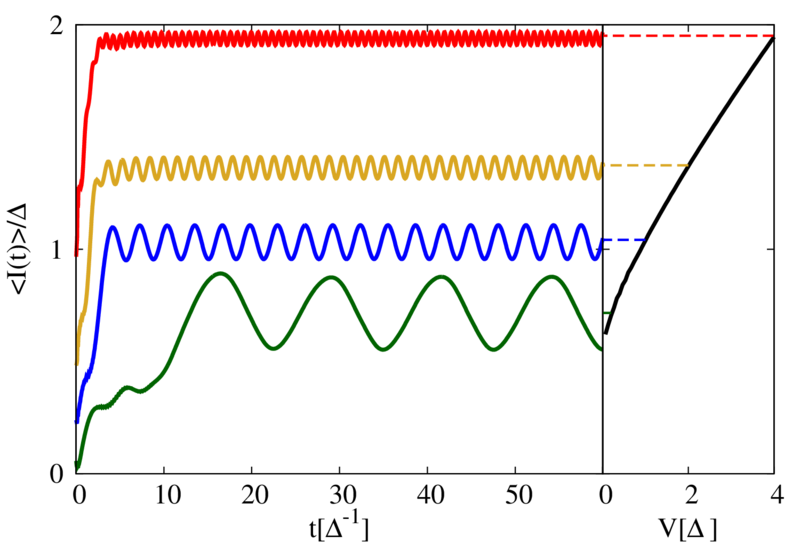}
\caption{Time evolution of the current of a biased voltage nanojunction. We show the convergence to the stationary current as a function of the time, for $V=4$, $2$, $1$ and $0.25\Delta$, from top to
bottom. We study the point contact regime, $\Gamma=10$ for a perfect transmitting junction ($\epsilon=0$ and $\Gamma_L=\Gamma_R=5$). In the right panel we show the convergence of the long time current 
\cite{Averin_PRL_1995,Cuevas_PRB_1996}.}
\label{V_behavior}
\end{figure}

\begin{figure}
\includegraphics[width=1\linewidth]{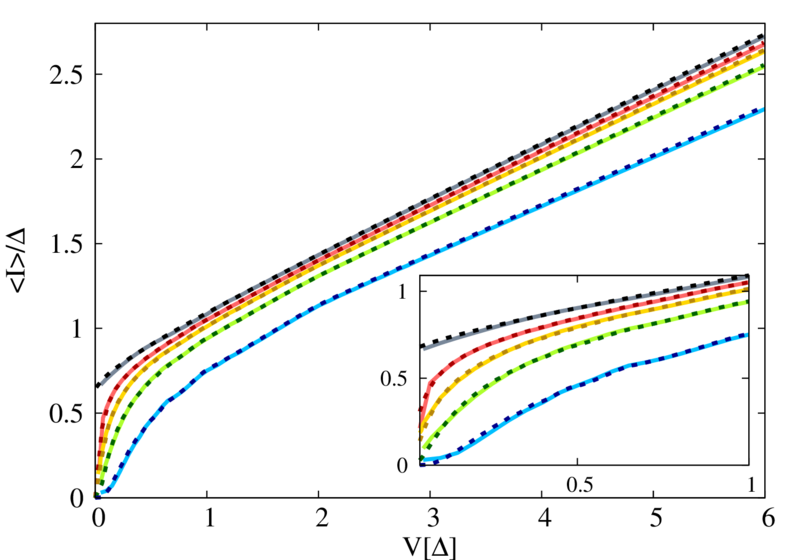}
\caption{Long time averaged current for different transmissions ($\tau=1$, $0.99$, $0.98$, $0.96$ and $0.9$, from top to bottom) in the QPC regime (we have chosen $\Gamma=60\Delta$), compared to the dc
stationary values (dashed lines) \cite{Averin_PRL_1995,Cuevas_PRB_1996}. Inset: zoom on the low bias limit.}
\label{I_G60}
\end{figure}

In Fig. \ref{V_behavior} we show the time evolution of the mean 
current for different bias voltages in the $\Gamma\gg\Delta$ regime. Differently from the phase-biased situation, the system relaxes to the stationary regime, with a relaxation time of the order of  
$t\sim \pi/V$. At longer times, the ac current oscillates around its mean value, 
represented in the right panel of Fig. \ref{V_behavior}. The oscillations with period $\pi/V$ correspond to the ac Josephson effect.
In Fig. \ref{I_G60} we show the long time averaged (dc) current in the QPC regime
for different transmission values. These results are in excellent agreement 
with the dc current obtained by standard stationary methods in Refs. 
\cite{Bratus_PRL1995,Averin_PRL_1995,Cuevas_PRB_1996}, showed as dashed lines in the figure. This agreement is poorer in the low bias regime $V\lesssim \Delta/10$, where the 
convergence time to reach the steady state becomes larger and the calculation becomes computationally more demanding.\\

\begin{figure}
\includegraphics[width=1\linewidth]{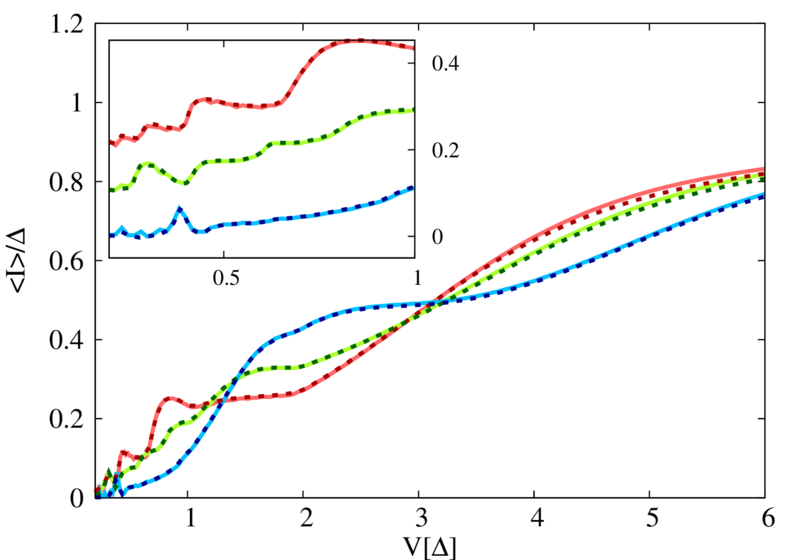}
\caption{Long time averaged current for different positions, $\epsilon=0$ (red), $0.5$ (green) and $1$ (blue) in units of $\Delta$  (corresponding to transmissions values 
$\tau=1$, $0.8$ and $0.5$, respectively) for the QD regime (We have chosen $\Gamma=\Delta$), compared to the stationary values (dashed lines) \cite{Yeyati_PRB_1997}. Inset: zoom on the low bias limit. 
Curves are shifted up for clarity.}
\label{I_G1}
\end{figure}

In Fig. \ref{I_G1} we show the dc current for a voltage biased junction in the QD regime. As in the QPC regime, we observe a remarkable agreement between the stationary calculation results 
\cite{Yeyati_PRB_1997,Johansson_PRB1999,Yeyati_PRL_2003} and the results obtained in this work in the long time limit. In the inset we show results for voltages smaller than the superconducting gap, exhibiting the expected 
subgap structure due to multiple Andreev reflections.\newline

\begin{figure}
\includegraphics[width=1\linewidth]{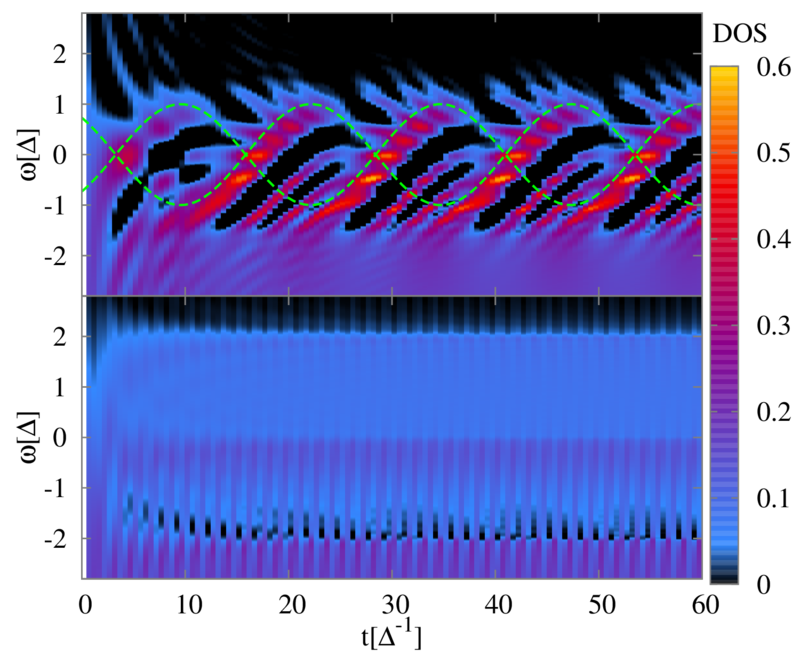}
\caption{Time evolution of the occupied DOS for a bias voltage nanojunction in the QPC regime. We consider the perfect transmitting situation with two bias voltage, $V=0.25$ (top panel) and 
$2\Delta$ (bottom panel). For a small voltage we observe an adiabatic evolution of the states (showed by the green dashed line), exchanging charge at their crossing points. 
This charge is relaxed to the continuum of 
states, generating quasiparticles which are able to decay in a time period. For a voltage bigger than the superconducting gap (bottom panel), the dynamics of the ABSs cannot be resolved in time. 
In this situation, the generated quasiparticles are not able to relax in a time period, leading to an almost constant density of states in the voltage window.}
\label{acJosephson_o}
\end{figure}

The convergence to the stationary situation is also illustrated in the time evolution of the occupied DOS in the QPC regime. In Fig. \ref{acJosephson_o} we show the
results for the case of a subgap voltage. Differently to the phase biased situation, the generated non-equilibrium quasi-particles relax when the ABSs approach the continuum of states.
After a few cycles, the system reaches the stationary condition with the states pumping charge from the lower to the upper continuum of states \cite{Yeyati_PRL_2003}. 
In addition to the two main features, which can be associated to the evolution of the ABSs, more structure appears as replicas (or satellites) of the ABSs, due to their non-adiabatic evolution.\newline

In the regime of $V>\Delta$, the evolution of the states becomes strongly non-adiabatic, being progressively difficult to resolve them in the DOS. In this regime, we observe a density of excited 
quasiparticles which is unable to relax in a time period. In the limit $V\gg\Delta$, we observe an almost homogeneous density of states, in the voltage window.\newline

\begin{figure}
\includegraphics[width=1\linewidth]{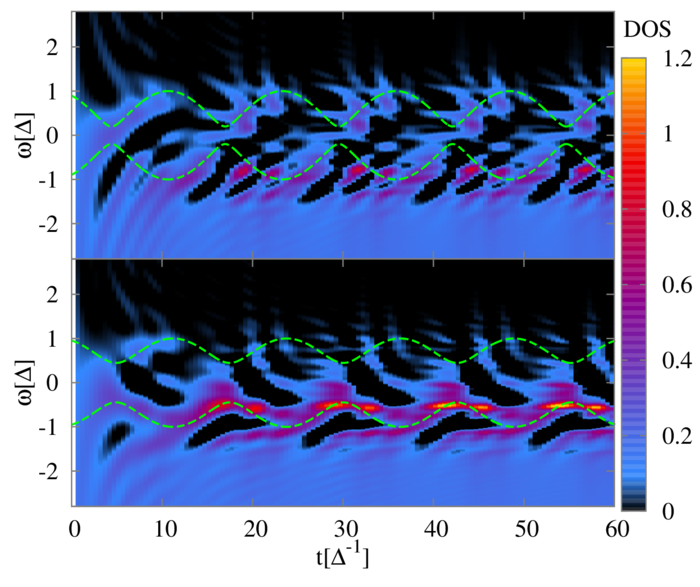}
\caption{Time evolution of the occupied DOS for a bias voltage nanojunction in the QPC regime, with transmissions $\tau=0.96$ (top panel) and $\tau=0.8$ (bottom panel), illustrating 
the convergence to the steady state independently from the value of the Andreev gap, $\Delta_A$.
The remaining parameters are the same as in the top panel of Fig. \ref{acJosephson_o}.}
\label{acJosephson_tau.8}
\end{figure}

For a non-perfect transmitting junction, an energy gap opens between the two ABSs which increases with decreasing $\tau$ as $\Delta_A=2\Delta\sqrt{1-\tau}$.
This situation was discussed in the stationary and low voltage regime in Refs. \cite{Averin_PRL_1995,Yeyati_PRL_2003}. In these works the authors demonstrated that the system evolves adiabatically except 
when $V\ll\Delta_A$
Landau-Zener transitions between the states, which happens with a probability $P=\exp[-\pi\Delta(1-\tau)/V]$. In Fig. \ref{acJosephson_tau.8} we show the time evolution of the occupied DOS 
for a non-perfect transmitting case with a voltage comparable (top panel) and much smaller (bottom panel) than the 
Andreev gap. In the first case, where the transition probability between the ABSs is $\sim0.5$, we observe some finite population of the upper ABS. In the second case, the transition
probability is negligible and the upper ABS 
remains almost unpopulated. Remarkably, we observe in both cases a convergence to the steady state, independently from
the initial conditions.\newline

\subsection{Full Counting statistics and dynamical Yang-Lee zeros}
\begin{figure}
\includegraphics[width=1\linewidth]{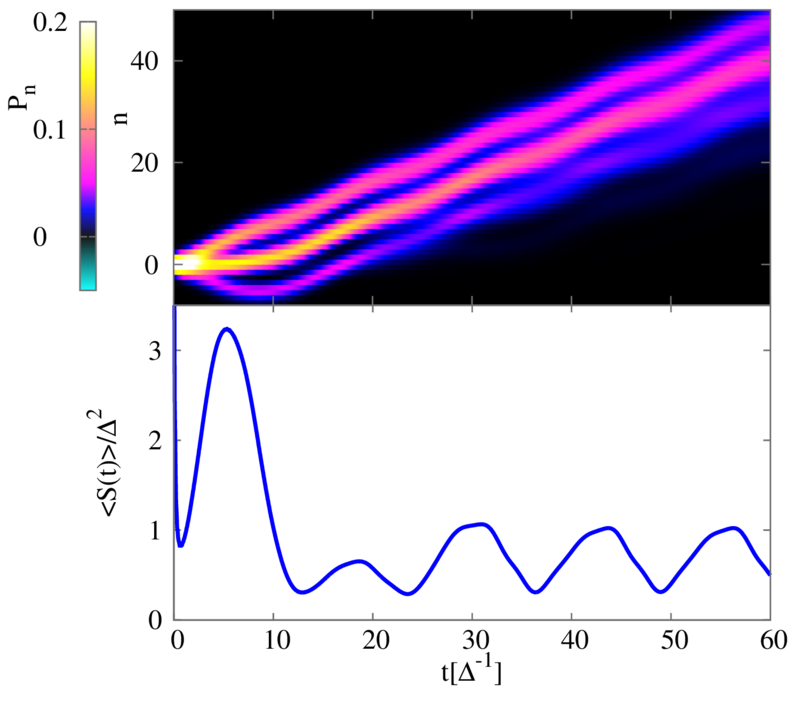}
\caption{Full counting statistics of a dc-biased junction For the same parameters as in the top panel of FIG. \ref{acJosephson_o}. In the top panel we show the time evolution of the probability 
after the contact formation. In the lower panel we show the evolution of the shot noise,
which exhibits the linear divergence of the shot noise, due to phase coexistence, which is relax in the time when the state reaches the continuum and quasiparticles are relax.}
\label{acJosephson_cumulants}
\end{figure}

In this subsection we present the FCS results for a voltage biased nanojunction after the sudden quench of the tunneling rates. In the upper panel of Fig. \ref{acJosephson_cumulants} we show the 
time evolution of the quasi-probabilities for the case $V<\Delta$. At very short times, smaller than the inverse of the Josephson frequency ($t\lesssim\pi/V$), three maxima are observed, i.e. a
signature of a phase coexistence between the three many body states described above. At longer times, the slopes of the three peaks become equal, reflecting the convergence
to the stationary regime characterized by the presence of a single quantum phase. The observation of these three maxima in the GF is related to the fact that the
probabilities are accumulated quantities, but it is no longer reflecting a coexistence between three phases at long times. Although not shown, for voltages $V\gtrsim\Delta$, the initially trapped 
quasiparticles are able to relax before the ABSs are fully developed, avoiding the short time phase coexistence, and exhibiting a single quasiprobability maximum evolving linearly in time.\newline

In the lower panel of Fig. \ref{acJosephson_cumulants} we show the time evolution of the current second cumulant, which can be related to the shot noise, for different bias voltages. At very short 
times, a linear increase in the noise is observed, consistent with the coexistence between the three phases. At longer times, when the phase coexistence disappears, the noise relaxes to the 
stationary situation. In the stationary regime, the shot noise exhibits an oscillatory behavior, where the maxima correspond to the subgap states approaching the gap edge leading to a maximum 
quasiparticle transfer.\newline

\begin{figure}
\includegraphics[width=1\linewidth]{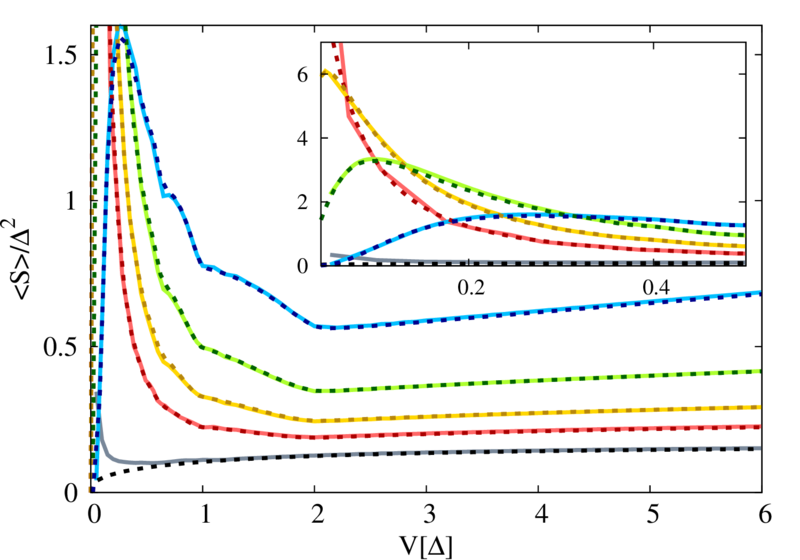}
\caption{Long time averaged noise for different transmissions ($\tau=1$, $0.99$, $0.98$, $0.96$ and $0.9$, from top to bottom) in the QPC regime (We have chosen $\Gamma=60\Delta$), compared to the 
stationary values (dashed lines) \cite{Cuevas_PRL_1999,Cuevas_PRL_2003,Cuevas_PRB_2004}. Inset: zoom on the low biased junction.}
\label{S_G60}
\end{figure}

The long time averaged shot noise behavior in the QPC regime is shown in Fig. \ref{S_G60}, comparing our numerical results (solid lines) with the expected stationary value (dashed lines)
\cite{Cuevas_PRL_1999,Cuevas_PRL_2003,Cuevas_PRB_2004}, for different transmission coefficients. As can be more clearly observed in the inset, where we show the shot noise in a enlarged scale
for subgap voltages, the agreement with the stationary results is quite remarkable, except for the extremely small voltages where the relaxation time becomes too long to be reached in our simulations. 
The effect of a finite $\Gamma$ value has also some influence in the deviation between both calculations observed in the limit $V\to0$ and $\tau\to1$, as the stationary calculation corresponds strictly 
to the $\Gamma\to\infty$ limit. As already discussed in Refs. \cite{Cuevas_PRL_1999,Cuevas_PRL_2003,Cuevas_PRB_2004}, the Fano factor $\left\langle S\right\rangle/\left\langle I\right\rangle$ diverges
in the $V\to0$ limit, reflecting the increase in the effective transmitted charge due to the multiple Andreev reflection processes of increasing order.\newline

\begin{figure}
\includegraphics[width=1\linewidth]{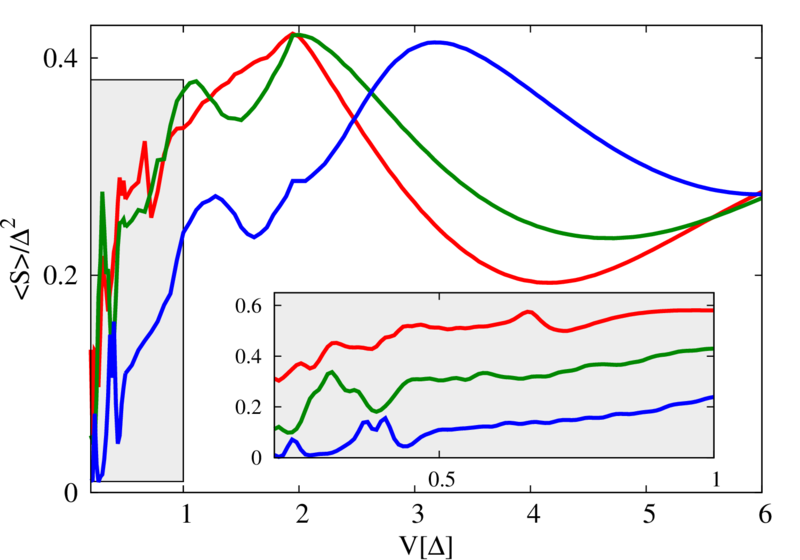}
\caption{Long time averaged noise for different positions, $\epsilon_0=0$, $0.5$ and $1$ (corresponding to transmissions $\tau=1$, $0.8$ and $0.5$, respectively) in the QD regime (We have chosen 
$\Gamma=\Delta$). Inset: zoom on the low biased junction where the curves are shifted up for clarity. The parameters are the same as in Fig. \ref{I_G1}.}
\label{S_G1}
\end{figure}

In Fig. \ref{S_G1}, we present the results for the long time averaged noise for the QD regime for three different level positions. In the inset we show the behavior for small voltages in an enlarged scale. 
To the best of our
knowledge, these results have not been reported before and can be relevant to describe recent experiments \cite{privateCom}. A detailed analysis of the observed features will be the subject of 
future work. It is worth remarking that, the time-resolved technique used in the present work allows us to
obtain results for the steady state properties in parameters regimes which could be inaccessible for other methods.\newline


\begin{figure}
\includegraphics[width=1\linewidth]{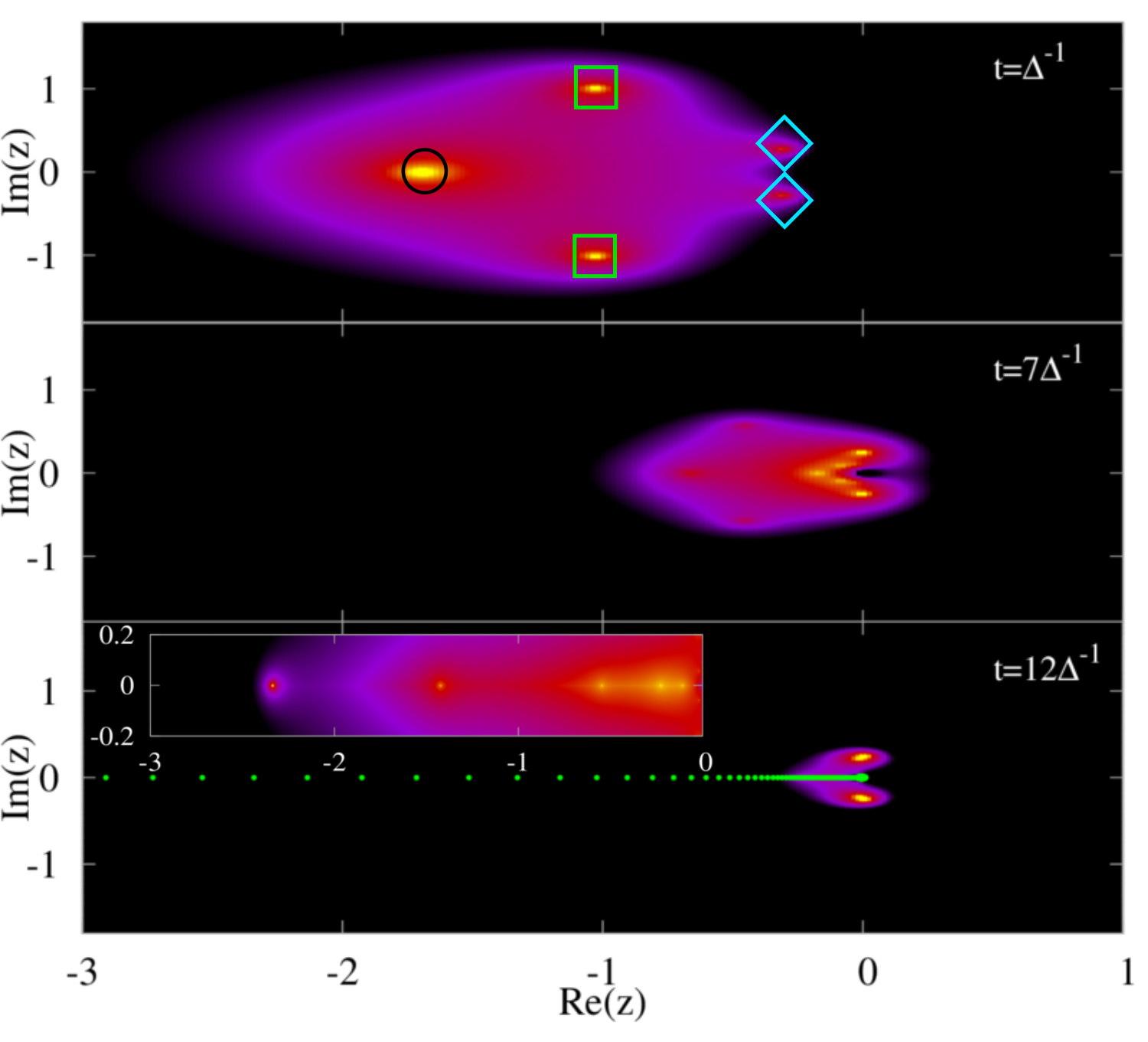}
\caption{DYLZs in the voltage biased situation for a perfect transmitting junction, $\Gamma=10$, $V=2\Delta$. We observe how the phase coexistence is broken when the ABSs approach the continuum of states.
In the top panel we show the result for the dominant zeros, marked as green squares and blue diamonds, as in the middle panel of Fig. \ref{poles:G2}. We observe an additional DYLZ related to the 
quasiparticle relaxation, marked with a black circle. For longer times, middle panel, their process become dominant and the zeros tend to converge to the negative real axis (bottom panel), showing
and small imaginary part close to $z=0$, which decreases with increasing $\Gamma$.
In the lower panel we show also the stationary result, computed using the  CGF described in Refs. \cite{Cuevas_PRL_2003,Cuevas_PRB_2004}, as green dots. The inset of the lower panel shows a zoom
close to the real axis, in order to show the convergence of the DYLZs to the negative real axis, showing a higher density close to $z=0$, in qualitative agreement to the stationary result.}
\label{poles_V}
\end{figure}

In Fig. \ref{poles_V} we show the evolution of the DYLZs for a voltage biased junction with $V>\Delta$. As in the phase driven situation, the zeros appear as complex conjugate
pairs for times of the order of the inverse of the superconducting gap (upper panel of Fig. \ref{poles_V}). There is a relaxation of the initially trapped quasiparticles in the ABSs when they 
approach the continuum at $t\sim \pi/V$. This relaxation manifests itself in the appearance of an additional DYLZ in the negative real axis, marked with a black circle, which is absent in the phase 
biased case (middle panel of Fig. \ref{poles:G2}). When this zero
becomes dominant (i.e. when it approaches the coordinates origin), the rest of the dynamical zeros approach the negative real axis (middle panel of Fig. \ref{poles_V}). 
At longer times we observe that the zeros tend to converge to the negative real axis, 
showing a small imaginary part close to $z=0$, which decreases with increasing $\Gamma$. In the inset of the lower panel, we show in detail in a different scale the convergence of the DYLZs to the
real axis, exhibiting a higher density close to $z=0$. This result is in qualitative agreement with the steady state zeros (shown as green dots in the lower panel of Fig. \ref{poles_V}), computed 
using the  CGF described in Refs. \cite{Cuevas_PRL_2003,Cuevas_PRB_2004} and shown as green dots in Fig. \ref{poles_V}.\newline

\begin{figure}
\includegraphics[width=1\linewidth]{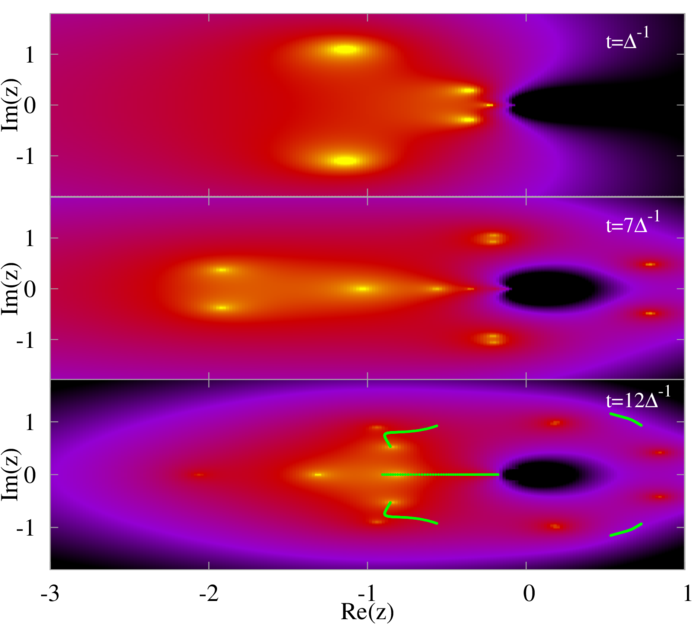}
\caption{Same as Fig. \ref{poles_V}, but with $V=0.25$. In the lower panel we compare the long time distribution of DYLZs with the stationary value, obtained from the CGF described in
 Refs. \cite{Cuevas_PRL_2003,Cuevas_PRB_2004}}
\label{poles_V.25}
\end{figure}

In Fig. \ref{poles_V.25} we show results for the DYLZs for a subgap bias. At short times (upper and middle panels) the zeros tend to converge to the unitary circle, which is a signature of a phase 
coexistence. Similarly to the case of voltages bigger than the superconducting gap, the DYLZs converge to their steady state, represented by the green lines in the lower panel of Fig. \ref{poles_V.25} 
and computed as above from the steady state results of Refs. \cite{Cuevas_PRL_2003,Cuevas_PRB_2004}. The number of stationary branches is related to the number of different multiple Andreev processes 
contributing to the charge transport through the system at the corresponding bias, roughly given by $2\Delta/V$.

\section{Initialization after a dc pulse}
\label{sec::dc_pulse}

\subsection{Current and ABS population}

\begin{figure}
\includegraphics[width=1\linewidth]{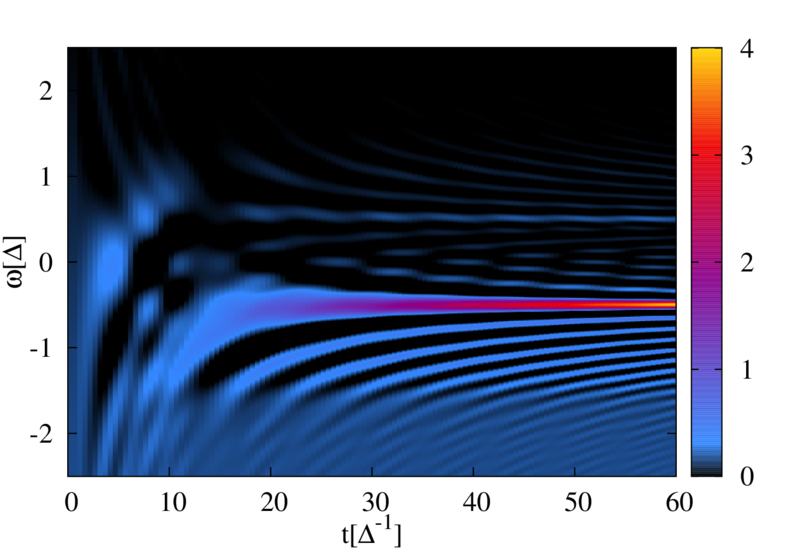}
\caption{Formation of the ABSs after a dc pulse of amplitude $V=2$, initializing the system in the ground state with a final phase difference $\phi=2$. The other parameters are the same as in Fig. 
\ref{V_behavior}.}
\label{acJosephson_o_initialized}
\end{figure}

The convergence to the stationary regime in the voltage biased case can be used to initialize the system in a given state, by applying short dc voltage pulses to the junction. This mechanism  
resembles the {\emph{antidote}} protocols proposed in Ref. \cite{Zgirski_PRL} to overcome quasiparticle poisoning. In Fig. \ref{acJosephson_o_initialized} we show the occupied DOS after a bias voltage 
sudden switch off, in the low 
voltage regime $V\lesssim\Delta$. We observe how the subgap states evolve towards their stationary values with an almost thermal equilibrium population after the pulse. There is still some small probability
of populating the upper state, given by the decaying quasiparticles from the upper continuum. It is important to note that the populated state is the one which has a positive dispersion relation (moving 
from negative to positive energies). It means that the upper ABS can be populated if the final phase is in the interval $\pi<\phi<2\pi$.\newline

\begin{figure}
\includegraphics[width=1\linewidth]{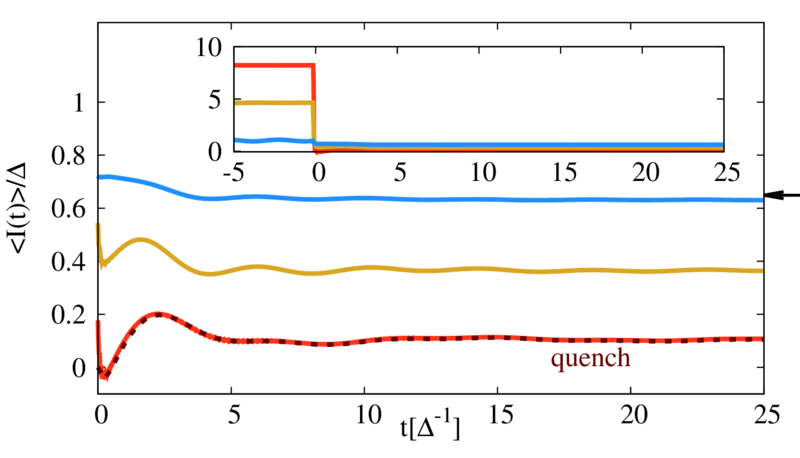}
\caption{Current after a bias voltage switch off at $t=0$ and a phase difference $\phi=2$. From top to bottom, $V=2$, $16$ and $100\Delta$ showing the transition from the convergence to the stationary 
current represented by the black arrow ($V\lesssim\Delta$) to the quench result represented by the discontinuous line ($V\gg\Delta$). In the inset we show the current evolution during and just after the 
dc pulse.}
\label{V_off}
\end{figure}

If a larger bias voltage is chosen ($V\gg\Delta$, $\Gamma$), a higher density of quasiparticles is generated and they are no longer relaxing to the expected thermal equilibrium situation after a
suden voltage switch off. These two opposite behaviors are illustrated by the
time evolution of the mean current in Fig. \ref{V_off}, showing a convergence to the thermal equilibrium situation for $V\lesssim\Delta$, and to memory-less quench dynamics for $V\gg\Delta$, $\Gamma$.\newline

\subsection{Full counting statistics and dynamical Yang-Lee zeros}

\begin{figure}
\includegraphics[width=1\linewidth]{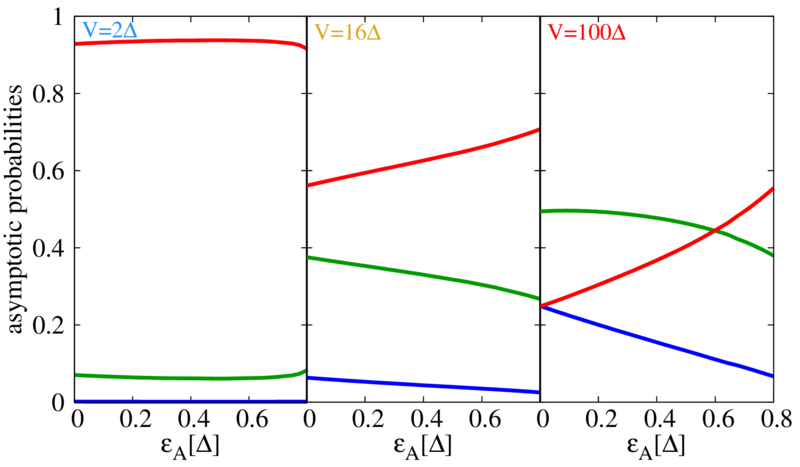}
\caption{Long time populations of the many body states after the voltage switch off. We show the same three situations represented in Fig. \ref{V_off}, which illustrates the transition from the 
equilibrium state (left) to the quench result (right) as the voltage is increased.}
\label{V_probs}
\end{figure}

The final state of the system can be better understood by analyzing the population of the many body states of the ABSs. This is illustrated in Fig. \ref{V_probs}. 
For small voltage pulses (left panel), we observe a probability of 
populating the ground state (red line) of more than $\sim90\%$. There is still some small probability of populating the odd state, given by the decaying quasiparticles in the upper continuum. Remarkably,
there is no probability of populating the excited even state, meaning that the probability of the charge to be excited from the lower of the upper state is negligible in this regime. When the voltage is
increased, we observe an evolution towards the universal quench result (right panel).\newline

\begin{figure}
\includegraphics[width=1\linewidth]{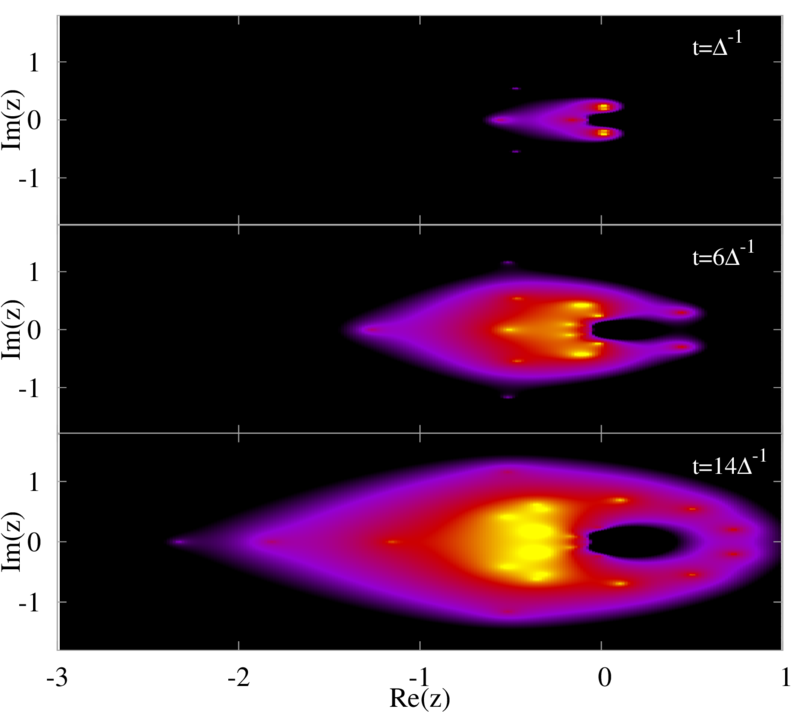}
\caption{Poles after a voltage pulse. We observe how gradually the zeros tend to describe a single circle, signature of coexistence between two phases (see left panel of Fig. \ref{V_probs}). Parameters:
$\Gamma=10$, $\tau=1$, amplitude of the pulse $V=2\Delta$, final phase $\phi=2$.}
\label{poles_switch}
\end{figure}

Finally, in Fig. \ref{poles_switch} we show the dynamical Yang-Lee zeros after a voltage quench. The system starts from the situation described in the lower panel of Fig. \ref{poles_V},
with most of the DYLZs accumulated at two points close to the origin. With increasing time, we observe the generation of a single circle, converging to the unitary one at long times. This
image is compatible with the two phases coexistence, as in the left panel of Fig. \ref{V_probs}.\newline

\section{Conclusions}
\label{sec::conclusions}
We have presented a comprehensive analysis of the transient dynamics associated with the formation of superconducting nanojunctions. We have shown how information on the mean transport quantities and on 
the many body states population can be extracted from the generating function of the FCS. In particular, we have shown how these properties can be related to the evolution of the 
zeros of the generating function. We have studied the quench dynamics both in the cases of phase and voltage biased nanojunctions. In the first case the system typically gets trapped in a metastable 
state which is dependent on the switch-on rate of the connection to the leads. There is also a sensitivity to initial conditions which is more pronounced in the QD regime.  In this case we have shown 
that either magnetic or non-magnetic metastable states can be produced. In the second case, the formation is accompanied by strong oscillations in the dot charge. Although the symmetrized current is not 
dependent on the initial conditions, their effect can be observed in the higher order cummulants and in the many body states population.
In the voltage biased case the system converges to the steady state, independently from the initial conditions. 
The results thus obtained for current and noise are in quantitative agreement with those obtained using conventional
stationary methods. We have also analyzed the possibility of coherently control the system state using a dc voltage drop. For small voltages ($V\lesssim\Delta$), the 
quasiparticles initially trapped in the system relax, and the system reaches the thermal equilibrium state. For large voltages, ($V\gg\Delta$), we recover the sudden connection result, providing a 
feasible experimental way to access the quench dynamics and to control the many body states population. We would like to remark that the method developed
in this work can be used to study more complex situations which cannot be accessed by conventional stationary approaches,
such as the one involving more terminals with non-commensurate applied bias 
\cite{Freyn_PRL_2011,Melin_PRB_2013,Pfeffer_PRB_2014} or the role of the interactions in the system.\newline

Finally it is worth noticing the connection between our work and the recent intense activity on the phenomenon of many body
localization (for recent reviews see \cite{Nandkishore,Vaseur_JSM}). Although these works are focused on closed many body systems, there is an analogy in the fact
that the system does not necessarily reach the thermal equilibrium state in the absence of coupling to an external bath, preserving the memory of the 
initial state. We thus believe that our approach could be of interest also in connection to this fundamental field of research.


\section{Acknowledgments}
We acknowledge discussions with W. Herrera,T. Jonckheere and J. P. Garrahan and financial support by Spanish MINECO through grant FIS2014-55486-P and the 
``Mar\'{\i}a de Maeztu'' Programme for Units of Excellence in R\&D (MDM-2014-0377). We also acknowledge
Santander Supercomputacion support group and  the Spanish Supercomputing Network (RES) for providing
access to the supercomputer Altamira at the Institute of Physics of Cantabria (IFCA-CSIC).

\bibliography{bibliography.bib}{}

\begin{thebibliography}{78}%
\makeatletter
\providecommand \@ifxundefined [1]{%
 \@ifx{#1\undefined}
}%
\providecommand \@ifnum [1]{%
 \ifnum #1\expandafter \@firstoftwo
 \else \expandafter \@secondoftwo
 \fi
}%
\providecommand \@ifx [1]{%
 \ifx #1\expandafter \@firstoftwo
 \else \expandafter \@secondoftwo
 \fi
}%
\providecommand \natexlab [1]{#1}%
\providecommand \enquote  [1]{``#1''}%
\providecommand \bibnamefont  [1]{#1}%
\providecommand \bibfnamefont [1]{#1}%
\providecommand \citenamefont [1]{#1}%
\providecommand \href@noop [0]{\@secondoftwo}%
\providecommand \href [0]{\begingroup \@sanitize@url \@href}%
\providecommand \@href[1]{\@@startlink{#1}\@@href}%
\providecommand \@@href[1]{\endgroup#1\@@endlink}%
\providecommand \@sanitize@url [0]{\catcode `\\12\catcode `\$12\catcode
  `\&12\catcode `\#12\catcode `\^12\catcode `\_12\catcode `\%12\relax}%
\providecommand \@@startlink[1]{}%
\providecommand \@@endlink[0]{}%
\providecommand \url  [0]{\begingroup\@sanitize@url \@url }%
\providecommand \@url [1]{\endgroup\@href {#1}{\urlprefix }}%
\providecommand \urlprefix  [0]{URL }%
\providecommand \Eprint [0]{\href }%
\providecommand \doibase [0]{http://dx.doi.org/}%
\providecommand \selectlanguage [0]{\@gobble}%
\providecommand \bibinfo  [0]{\@secondoftwo}%
\providecommand \bibfield  [0]{\@secondoftwo}%
\providecommand \translation [1]{[#1]}%
\providecommand \BibitemOpen [0]{}%
\providecommand \bibitemStop [0]{}%
\providecommand \bibitemNoStop [0]{.\EOS\space}%
\providecommand \EOS [0]{\spacefactor3000\relax}%
\providecommand \BibitemShut  [1]{\csname bibitem#1\endcsname}%
\let\auto@bib@innerbib\@empty
\bibitem [{\citenamefont {Devoret}\ and\ \citenamefont
  {Schoelkopf}(2013)}]{Devoret_Science}%
  \BibitemOpen
  \bibfield  {author} {\bibinfo {author} {\bibfnamefont {M.~H.}\ \bibnamefont
  {Devoret}}\ and\ \bibinfo {author} {\bibfnamefont {R.~J.}\ \bibnamefont
  {Schoelkopf}},\ }\bibfield  {title} {\enquote {\bibinfo {title}
  {Superconducting circuits for quantum information: An outlook},}\ }\href
  {\doibase 10.1126/science.1231930} {\bibfield  {journal} {\bibinfo  {journal}
  {Science}\ }\textbf {\bibinfo {volume} {339}},\ \bibinfo {pages} {1169--1174}
  (\bibinfo {year} {2013})}\BibitemShut {NoStop}%
\bibitem [{\citenamefont {Alicea}(2012)}]{Alicea_RPP}%
  \BibitemOpen
  \bibfield  {author} {\bibinfo {author} {\bibfnamefont {Jason}\ \bibnamefont
  {Alicea}},\ }\bibfield  {title} {\enquote {\bibinfo {title} {New directions
  in the pursuit of majorana fermions in solid state systems},}\ }\href
  {http://stacks.iop.org/0034-4885/75/i=7/a=076501} {\bibfield  {journal}
  {\bibinfo  {journal} {Reports on Progress in Physics}\ }\textbf {\bibinfo
  {volume} {75}},\ \bibinfo {pages} {076501} (\bibinfo {year}
  {2012})}\BibitemShut {NoStop}%
\bibitem [{\citenamefont {Beenakker}(2015)}]{Beenakker_RMP}%
  \BibitemOpen
  \bibfield  {author} {\bibinfo {author} {\bibfnamefont {C.~W.~J.}\
  \bibnamefont {Beenakker}},\ }\bibfield  {title} {\enquote {\bibinfo {title}
  {Random-matrix theory of majorana fermions and topological
  superconductors},}\ }\href {\doibase 10.1103/RevModPhys.87.1037} {\bibfield
  {journal} {\bibinfo  {journal} {Rev. Mod. Phys.}\ }\textbf {\bibinfo {volume}
  {87}},\ \bibinfo {pages} {1037--1066} (\bibinfo {year} {2015})}\BibitemShut
  {NoStop}%
\bibitem [{\citenamefont {Plugge}\ \emph {et~al.}(2017)\citenamefont {Plugge},
  \citenamefont {Rasmussen}, \citenamefont {Egger},\ and\ \citenamefont
  {Flensberg}}]{Flensberg_NJP2017}%
  \BibitemOpen
  \bibfield  {author} {\bibinfo {author} {\bibfnamefont {Stephan}\ \bibnamefont
  {Plugge}}, \bibinfo {author} {\bibfnamefont {Asbjørn}\ \bibnamefont
  {Rasmussen}}, \bibinfo {author} {\bibfnamefont {Reinhold}\ \bibnamefont
  {Egger}}, \ and\ \bibinfo {author} {\bibfnamefont {Karsten}\ \bibnamefont
  {Flensberg}},\ }\bibfield  {title} {\enquote {\bibinfo {title} {Majorana box
  qubits},}\ }\href {http://stacks.iop.org/1367-2630/19/i=1/a=012001}
  {\bibfield  {journal} {\bibinfo  {journal} {New Journal of Physics}\ }\textbf
  {\bibinfo {volume} {19}},\ \bibinfo {pages} {012001} (\bibinfo {year}
  {2017})}\BibitemShut {NoStop}%
\bibitem [{\citenamefont {Martinis}\ \emph {et~al.}(2009)\citenamefont
  {Martinis}, \citenamefont {Ansmann},\ and\ \citenamefont
  {Aumentado}}]{Martinis_PRL2009}%
  \BibitemOpen
  \bibfield  {author} {\bibinfo {author} {\bibfnamefont {John~M.}\ \bibnamefont
  {Martinis}}, \bibinfo {author} {\bibfnamefont {M.}~\bibnamefont {Ansmann}}, \
  and\ \bibinfo {author} {\bibfnamefont {J.}~\bibnamefont {Aumentado}},\
  }\bibfield  {title} {\enquote {\bibinfo {title} {Energy decay in
  superconducting josephson-junction qubits from nonequilibrium quasiparticle
  excitations},}\ }\href {\doibase 10.1103/PhysRevLett.103.097002} {\bibfield
  {journal} {\bibinfo  {journal} {Phys. Rev. Lett.}\ }\textbf {\bibinfo
  {volume} {103}},\ \bibinfo {pages} {097002} (\bibinfo {year}
  {2009})}\BibitemShut {NoStop}%
\bibitem [{\citenamefont {Catelani}\ \emph {et~al.}(2011)\citenamefont
  {Catelani}, \citenamefont {Schoelkopf}, \citenamefont {Devoret},\ and\
  \citenamefont {Glazman}}]{Catelani_PRB2011}%
  \BibitemOpen
  \bibfield  {author} {\bibinfo {author} {\bibfnamefont {G.}~\bibnamefont
  {Catelani}}, \bibinfo {author} {\bibfnamefont {R.~J.}\ \bibnamefont
  {Schoelkopf}}, \bibinfo {author} {\bibfnamefont {M.~H.}\ \bibnamefont
  {Devoret}}, \ and\ \bibinfo {author} {\bibfnamefont {L.~I.}\ \bibnamefont
  {Glazman}},\ }\bibfield  {title} {\enquote {\bibinfo {title} {Relaxation and
  frequency shifts induced by quasiparticles in superconducting qubits},}\
  }\href {\doibase 10.1103/PhysRevB.84.064517} {\bibfield  {journal} {\bibinfo
  {journal} {Phys. Rev. B}\ }\textbf {\bibinfo {volume} {84}},\ \bibinfo
  {pages} {064517} (\bibinfo {year} {2011})}\BibitemShut {NoStop}%
\bibitem [{\citenamefont {Rist{\`e}}\ \emph {et~al.}(2013)\citenamefont
  {Rist{\`e}}, \citenamefont {Bultink}, \citenamefont {Tiggelman},
  \citenamefont {Schouten}, \citenamefont {Lehnert},\ and\ \citenamefont
  {DiCarlo}}]{Riste_NatCom2013}%
  \BibitemOpen
  \bibfield  {author} {\bibinfo {author} {\bibfnamefont {D.}~\bibnamefont
  {Rist{\`e}}}, \bibinfo {author} {\bibfnamefont {C.~C.}\ \bibnamefont
  {Bultink}}, \bibinfo {author} {\bibfnamefont {M.~J.}\ \bibnamefont
  {Tiggelman}}, \bibinfo {author} {\bibfnamefont {R.~N.}\ \bibnamefont
  {Schouten}}, \bibinfo {author} {\bibfnamefont {K.~W.}\ \bibnamefont
  {Lehnert}}, \ and\ \bibinfo {author} {\bibfnamefont {L.}~\bibnamefont
  {DiCarlo}},\ }\bibfield  {title} {\enquote {\bibinfo {title} {Millisecond
  charge-parity fluctuations and induced decoherence in a superconducting
  transmon qubit},}\ }\href {\doibase 10.1038/ncomms2936} {\bibfield  {journal}
  {\bibinfo  {journal} {Nat. Commun.}\ }\textbf {\bibinfo {volume} {4}},\
  \bibinfo {pages} {1913} (\bibinfo {year} {2013})}\BibitemShut {NoStop}%
\bibitem [{\citenamefont {Avriller}\ and\ \citenamefont
  {Pistolesi}(2015)}]{Avriller_PRL2015}%
  \BibitemOpen
  \bibfield  {author} {\bibinfo {author} {\bibfnamefont {R.}~\bibnamefont
  {Avriller}}\ and\ \bibinfo {author} {\bibfnamefont {F.}~\bibnamefont
  {Pistolesi}},\ }\bibfield  {title} {\enquote {\bibinfo {title} {Andreev
  bound-state dynamics in quantum-dot josephson junctions: A washing out of the
  $0\text{\ensuremath{-}}\ensuremath{\pi}$ transition},}\ }\href {\doibase
  10.1103/PhysRevLett.114.037003} {\bibfield  {journal} {\bibinfo  {journal}
  {Phys. Rev. Lett.}\ }\textbf {\bibinfo {volume} {114}},\ \bibinfo {pages}
  {037003} (\bibinfo {year} {2015})}\BibitemShut {NoStop}%
\bibitem [{\citenamefont {Rainis}\ and\ \citenamefont
  {Loss}(2012)}]{Rainis_PRB}%
  \BibitemOpen
  \bibfield  {author} {\bibinfo {author} {\bibfnamefont {Diego}\ \bibnamefont
  {Rainis}}\ and\ \bibinfo {author} {\bibfnamefont {Daniel}\ \bibnamefont
  {Loss}},\ }\bibfield  {title} {\enquote {\bibinfo {title} {Majorana qubit
  decoherence by quasiparticle poisoning},}\ }\href {\doibase
  10.1103/PhysRevB.85.174533} {\bibfield  {journal} {\bibinfo  {journal} {Phys.
  Rev. B}\ }\textbf {\bibinfo {volume} {85}},\ \bibinfo {pages} {174533}
  (\bibinfo {year} {2012})}\BibitemShut {NoStop}%
\bibitem [{\citenamefont {Colbert}\ and\ \citenamefont
  {Lee}(2014)}]{Colbert_PRB2014}%
  \BibitemOpen
  \bibfield  {author} {\bibinfo {author} {\bibfnamefont {Jacob~R.}\
  \bibnamefont {Colbert}}\ and\ \bibinfo {author} {\bibfnamefont {Patrick~A.}\
  \bibnamefont {Lee}},\ }\bibfield  {title} {\enquote {\bibinfo {title}
  {Proposal to measure the quasiparticle poisoning time of majorana bound
  states},}\ }\href {\doibase 10.1103/PhysRevB.89.140505} {\bibfield  {journal}
  {\bibinfo  {journal} {Phys. Rev. B}\ }\textbf {\bibinfo {volume} {89}},\
  \bibinfo {pages} {140505} (\bibinfo {year} {2014})}\BibitemShut {NoStop}%
\bibitem [{\citenamefont {Bespalov}\ \emph {et~al.}(2016)\citenamefont
  {Bespalov}, \citenamefont {Houzet}, \citenamefont {Meyer},\ and\
  \citenamefont {Nazarov}}]{Bespalov_PRL2016}%
  \BibitemOpen
  \bibfield  {author} {\bibinfo {author} {\bibfnamefont {Anton}\ \bibnamefont
  {Bespalov}}, \bibinfo {author} {\bibfnamefont {Manuel}\ \bibnamefont
  {Houzet}}, \bibinfo {author} {\bibfnamefont {Julia~S.}\ \bibnamefont
  {Meyer}}, \ and\ \bibinfo {author} {\bibfnamefont {Yuli~V.}\ \bibnamefont
  {Nazarov}},\ }\bibfield  {title} {\enquote {\bibinfo {title} {Theoretical
  model to explain excess of quasiparticles in superconductors},}\ }\href
  {\doibase 10.1103/PhysRevLett.117.117002} {\bibfield  {journal} {\bibinfo
  {journal} {Phys. Rev. Lett.}\ }\textbf {\bibinfo {volume} {117}},\ \bibinfo
  {pages} {117002} (\bibinfo {year} {2016})}\BibitemShut {NoStop}%
\bibitem [{\citenamefont {Albrecht}\ \emph {et~al.}(2017)\citenamefont
  {Albrecht}, \citenamefont {Hansen}, \citenamefont {Higginbotham},
  \citenamefont {Kuemmeth}, \citenamefont {Jespersen}, \citenamefont
  {Nyg\aa{}rd}, \citenamefont {Krogstrup}, \citenamefont {Danon}, \citenamefont
  {Flensberg},\ and\ \citenamefont {Marcus}}]{Flensberg_PRL2017}%
  \BibitemOpen
  \bibfield  {author} {\bibinfo {author} {\bibfnamefont {S.~M.}\ \bibnamefont
  {Albrecht}}, \bibinfo {author} {\bibfnamefont {E.~B.}\ \bibnamefont
  {Hansen}}, \bibinfo {author} {\bibfnamefont {A.~P.}\ \bibnamefont
  {Higginbotham}}, \bibinfo {author} {\bibfnamefont {F.}~\bibnamefont
  {Kuemmeth}}, \bibinfo {author} {\bibfnamefont {T.~S.}\ \bibnamefont
  {Jespersen}}, \bibinfo {author} {\bibfnamefont {J.}~\bibnamefont
  {Nyg\aa{}rd}}, \bibinfo {author} {\bibfnamefont {P.}~\bibnamefont
  {Krogstrup}}, \bibinfo {author} {\bibfnamefont {J.}~\bibnamefont {Danon}},
  \bibinfo {author} {\bibfnamefont {K.}~\bibnamefont {Flensberg}}, \ and\
  \bibinfo {author} {\bibfnamefont {C.~M.}\ \bibnamefont {Marcus}},\ }\bibfield
   {title} {\enquote {\bibinfo {title} {Transport signatures of quasiparticle
  poisoning in a majorana island},}\ }\href {\doibase
  10.1103/PhysRevLett.118.137701} {\bibfield  {journal} {\bibinfo  {journal}
  {Phys. Rev. Lett.}\ }\textbf {\bibinfo {volume} {118}},\ \bibinfo {pages}
  {137701} (\bibinfo {year} {2017})}\BibitemShut {NoStop}%
\bibitem [{\citenamefont {Zgirski}\ \emph {et~al.}(2011)\citenamefont
  {Zgirski}, \citenamefont {Bretheau}, \citenamefont {Le~Masne}, \citenamefont
  {Pothier}, \citenamefont {Esteve},\ and\ \citenamefont
  {Urbina}}]{Zgirski_PRL}%
  \BibitemOpen
  \bibfield  {author} {\bibinfo {author} {\bibfnamefont {M.}~\bibnamefont
  {Zgirski}}, \bibinfo {author} {\bibfnamefont {L.}~\bibnamefont {Bretheau}},
  \bibinfo {author} {\bibfnamefont {Q.}~\bibnamefont {Le~Masne}}, \bibinfo
  {author} {\bibfnamefont {H.}~\bibnamefont {Pothier}}, \bibinfo {author}
  {\bibfnamefont {D.}~\bibnamefont {Esteve}}, \ and\ \bibinfo {author}
  {\bibfnamefont {C.}~\bibnamefont {Urbina}},\ }\bibfield  {title} {\enquote
  {\bibinfo {title} {Evidence for long-lived quasiparticles trapped in
  superconducting point contacts},}\ }\href {\doibase
  10.1103/PhysRevLett.106.257003} {\bibfield  {journal} {\bibinfo  {journal}
  {Phys. Rev. Lett.}\ }\textbf {\bibinfo {volume} {106}},\ \bibinfo {pages}
  {257003} (\bibinfo {year} {2011})}\BibitemShut {NoStop}%
\bibitem [{\citenamefont {Olivares}\ \emph {et~al.}(2014)\citenamefont
  {Olivares}, \citenamefont {Yeyati}, \citenamefont {Bretheau}, \citenamefont
  {Girit}, \citenamefont {Pothier},\ and\ \citenamefont
  {Urbina}}]{Olivares_PRB2014}%
  \BibitemOpen
  \bibfield  {author} {\bibinfo {author} {\bibfnamefont {D.~G.}\ \bibnamefont
  {Olivares}}, \bibinfo {author} {\bibfnamefont {A.~L.}\ \bibnamefont
  {Yeyati}}, \bibinfo {author} {\bibfnamefont {L.}~\bibnamefont {Bretheau}},
  \bibinfo {author} {\bibfnamefont {{\c C}.~{\"O}.}\ \bibnamefont {Girit}},
  \bibinfo {author} {\bibfnamefont {H.}~\bibnamefont {Pothier}}, \ and\
  \bibinfo {author} {\bibfnamefont {C.}~\bibnamefont {Urbina}},\ }\bibfield
  {title} {\enquote {\bibinfo {title} {Dynamics of quasiparticle trapping in
  andreev levels},}\ }\href {\doibase 10.1103/PhysRevB.89.104504} {\bibfield
  {journal} {\bibinfo  {journal} {Phys. Rev. B}\ }\textbf {\bibinfo {volume}
  {89}},\ \bibinfo {pages} {104504} (\bibinfo {year} {2014})}\BibitemShut
  {NoStop}%
\bibitem [{\citenamefont {Padurariu}\ and\ \citenamefont
  {Nazarov}(2012)}]{Padurario_EPL}%
  \BibitemOpen
  \bibfield  {author} {\bibinfo {author} {\bibfnamefont {C.}~\bibnamefont
  {Padurariu}}\ and\ \bibinfo {author} {\bibfnamefont {Yu.~V.}\ \bibnamefont
  {Nazarov}},\ }\bibfield  {title} {\enquote {\bibinfo {title} {Spin blockade
  qubit in a superconducting junction},}\ }\href
  {http://stacks.iop.org/0295-5075/100/i=5/a=57006} {\bibfield  {journal}
  {\bibinfo  {journal} {EPL (Europhysics Letters)}\ }\textbf {\bibinfo {volume}
  {100}},\ \bibinfo {pages} {57006} (\bibinfo {year} {2012})}\BibitemShut
  {NoStop}%
\bibitem [{\citenamefont {Nazarov}\ and\ \citenamefont
  {Yaroslav}(2009)}]{Nazarov_book}%
  \BibitemOpen
  \bibfield  {author} {\bibinfo {author} {\bibfnamefont {Yuli~V}\ \bibnamefont
  {Nazarov}}\ and\ \bibinfo {author} {\bibfnamefont {Blanter~M}\ \bibnamefont
  {Yaroslav}},\ }\href@noop {} {\emph {\bibinfo {title} {{Quantum transport:
  introduction to nanoscience}}}}\ (\bibinfo  {publisher} {Cambridge Univ.
  Press},\ \bibinfo {address} {Cambridge},\ \bibinfo {year} {2009})\BibitemShut
  {NoStop}%
\bibitem [{\citenamefont {F{\`e}ve}\ \emph {et~al.}(2007)\citenamefont
  {F{\`e}ve}, \citenamefont {Mah{\'e}}, \citenamefont {Berroir}, \citenamefont
  {Kontos}, \citenamefont {Pla{\c c}ais}, \citenamefont {Glattli},
  \citenamefont {Cavanna}, \citenamefont {Etienne},\ and\ \citenamefont
  {Jin}}]{FeveScience}%
  \BibitemOpen
  \bibfield  {author} {\bibinfo {author} {\bibfnamefont {G.}~\bibnamefont
  {F{\`e}ve}}, \bibinfo {author} {\bibfnamefont {A.}~\bibnamefont {Mah{\'e}}},
  \bibinfo {author} {\bibfnamefont {J.-M.}\ \bibnamefont {Berroir}}, \bibinfo
  {author} {\bibfnamefont {T.}~\bibnamefont {Kontos}}, \bibinfo {author}
  {\bibfnamefont {B.}~\bibnamefont {Pla{\c c}ais}}, \bibinfo {author}
  {\bibfnamefont {D.~C.}\ \bibnamefont {Glattli}}, \bibinfo {author}
  {\bibfnamefont {A.}~\bibnamefont {Cavanna}}, \bibinfo {author} {\bibfnamefont
  {B.}~\bibnamefont {Etienne}}, \ and\ \bibinfo {author} {\bibfnamefont
  {Y.}~\bibnamefont {Jin}},\ }\bibfield  {title} {\enquote {\bibinfo {title}
  {An on-demand coherent single-electron source},}\ }\href {\doibase
  10.1126/science.1141243} {\bibfield  {journal} {\bibinfo  {journal}
  {Science}\ }\textbf {\bibinfo {volume} {316}},\ \bibinfo {pages} {1169--1172}
  (\bibinfo {year} {2007})}\BibitemShut {NoStop}%
\bibitem [{\citenamefont {Bocquillon}\ \emph {et~al.}(2013)\citenamefont
  {Bocquillon}, \citenamefont {Freulon}, \citenamefont {Berroir}, \citenamefont
  {Degiovanni}, \citenamefont {Pla{\c c}ais}, \citenamefont {Cavanna},
  \citenamefont {Jin},\ and\ \citenamefont {F{\`e}ve}}]{BocquillonScience}%
  \BibitemOpen
  \bibfield  {author} {\bibinfo {author} {\bibfnamefont {E.}~\bibnamefont
  {Bocquillon}}, \bibinfo {author} {\bibfnamefont {V.}~\bibnamefont {Freulon}},
  \bibinfo {author} {\bibfnamefont {J.-M}\ \bibnamefont {Berroir}}, \bibinfo
  {author} {\bibfnamefont {P.}~\bibnamefont {Degiovanni}}, \bibinfo {author}
  {\bibfnamefont {B.}~\bibnamefont {Pla{\c c}ais}}, \bibinfo {author}
  {\bibfnamefont {A.}~\bibnamefont {Cavanna}}, \bibinfo {author} {\bibfnamefont
  {Y.}~\bibnamefont {Jin}}, \ and\ \bibinfo {author} {\bibfnamefont
  {G.}~\bibnamefont {F{\`e}ve}},\ }\bibfield  {title} {\enquote {\bibinfo
  {title} {Coherence and indistinguishability of single electrons emitted by
  independent sources},}\ }\href {\doibase 10.1126/science.1232572} {\bibfield
  {journal} {\bibinfo  {journal} {Science}\ }\textbf {\bibinfo {volume}
  {339}},\ \bibinfo {pages} {1054--1057} (\bibinfo {year} {2013})}\BibitemShut
  {NoStop}%
\bibitem [{\citenamefont {Dubois}\ \emph {et~al.}(2013)\citenamefont {Dubois},
  \citenamefont {Jullien}, \citenamefont {Portier}, \citenamefont {Roche},
  \citenamefont {Cavanna}, \citenamefont {Jin}, \citenamefont {Wegscheider},
  \citenamefont {Roulleau},\ and\ \citenamefont {Glattli}}]{DuboisNature}%
  \BibitemOpen
  \bibfield  {author} {\bibinfo {author} {\bibfnamefont {J.}~\bibnamefont
  {Dubois}}, \bibinfo {author} {\bibfnamefont {T.}~\bibnamefont {Jullien}},
  \bibinfo {author} {\bibfnamefont {F.}~\bibnamefont {Portier}}, \bibinfo
  {author} {\bibfnamefont {P.}~\bibnamefont {Roche}}, \bibinfo {author}
  {\bibfnamefont {A.}~\bibnamefont {Cavanna}}, \bibinfo {author} {\bibfnamefont
  {Y.}~\bibnamefont {Jin}}, \bibinfo {author} {\bibfnamefont {W.}~\bibnamefont
  {Wegscheider}}, \bibinfo {author} {\bibfnamefont {P.}~\bibnamefont
  {Roulleau}}, \ and\ \bibinfo {author} {\bibfnamefont {D.~C.}\ \bibnamefont
  {Glattli}},\ }\bibfield  {title} {\enquote {\bibinfo {title}
  {Minimal-excitation states for electron quantum optics using levitons},}\
  }\href {http://dx.doi.org/10.1038/nature12713} {\bibfield  {journal}
  {\bibinfo  {journal} {Nature (London)}\ }\textbf {\bibinfo {volume} {502}},\
  \bibinfo {pages} {659--663} (\bibinfo {year} {2013})}\BibitemShut {NoStop}%
\bibitem [{\citenamefont {Garrahan}\ and\ \citenamefont
  {Lesanovsky}(2010)}]{Garrahan_PRL2010}%
  \BibitemOpen
  \bibfield  {author} {\bibinfo {author} {\bibfnamefont {Juan~P.}\ \bibnamefont
  {Garrahan}}\ and\ \bibinfo {author} {\bibfnamefont {Igor}\ \bibnamefont
  {Lesanovsky}},\ }\bibfield  {title} {\enquote {\bibinfo {title}
  {Thermodynamics of quantum jump trajectories},}\ }\href {\doibase
  10.1103/PhysRevLett.104.160601} {\bibfield  {journal} {\bibinfo  {journal}
  {Phys. Rev. Lett.}\ }\textbf {\bibinfo {volume} {104}},\ \bibinfo {pages}
  {160601} (\bibinfo {year} {2010})}\BibitemShut {NoStop}%
\bibitem [{\citenamefont {Heyl}\ \emph {et~al.}(2013)\citenamefont {Heyl},
  \citenamefont {Polkovnikov},\ and\ \citenamefont {Kehrein}}]{Heyl_PRL2013}%
  \BibitemOpen
  \bibfield  {author} {\bibinfo {author} {\bibfnamefont {M.}~\bibnamefont
  {Heyl}}, \bibinfo {author} {\bibfnamefont {A.}~\bibnamefont {Polkovnikov}}, \
  and\ \bibinfo {author} {\bibfnamefont {S.}~\bibnamefont {Kehrein}},\
  }\bibfield  {title} {\enquote {\bibinfo {title} {Dynamical quantum phase
  transitions in the transverse-field ising model},}\ }\href {\doibase
  10.1103/PhysRevLett.110.135704} {\bibfield  {journal} {\bibinfo  {journal}
  {Phys. Rev. Lett.}\ }\textbf {\bibinfo {volume} {110}},\ \bibinfo {pages}
  {135704} (\bibinfo {year} {2013})}\BibitemShut {NoStop}%
\bibitem [{\citenamefont {Karrasch}\ and\ \citenamefont
  {Schuricht}(2013)}]{Karrasch_PRB2013}%
  \BibitemOpen
  \bibfield  {author} {\bibinfo {author} {\bibfnamefont {C.}~\bibnamefont
  {Karrasch}}\ and\ \bibinfo {author} {\bibfnamefont {D.}~\bibnamefont
  {Schuricht}},\ }\bibfield  {title} {\enquote {\bibinfo {title} {Dynamical
  phase transitions after quenches in nonintegrable models},}\ }\href {\doibase
  10.1103/PhysRevB.87.195104} {\bibfield  {journal} {\bibinfo  {journal} {Phys.
  Rev. B}\ }\textbf {\bibinfo {volume} {87}},\ \bibinfo {pages} {195104}
  (\bibinfo {year} {2013})}\BibitemShut {NoStop}%
\bibitem [{\citenamefont {Hickey}\ \emph
  {et~al.}(2013{\natexlab{a}})\citenamefont {Hickey}, \citenamefont {Genway},
  \citenamefont {Lesanovsky},\ and\ \citenamefont
  {Garrahan}}]{Garrahan_PRB2013}%
  \BibitemOpen
  \bibfield  {author} {\bibinfo {author} {\bibfnamefont {James~M.}\
  \bibnamefont {Hickey}}, \bibinfo {author} {\bibfnamefont {Sam}\ \bibnamefont
  {Genway}}, \bibinfo {author} {\bibfnamefont {Igor}\ \bibnamefont
  {Lesanovsky}}, \ and\ \bibinfo {author} {\bibfnamefont {Juan~P.}\
  \bibnamefont {Garrahan}},\ }\bibfield  {title} {\enquote {\bibinfo {title}
  {Time-integrated observables as order parameters for full counting statistics
  transitions in closed quantum systems},}\ }\href {\doibase
  10.1103/PhysRevB.87.184303} {\bibfield  {journal} {\bibinfo  {journal} {Phys.
  Rev. B}\ }\textbf {\bibinfo {volume} {87}},\ \bibinfo {pages} {184303}
  (\bibinfo {year} {2013}{\natexlab{a}})}\BibitemShut {NoStop}%
\bibitem [{\citenamefont {Yang}\ and\ \citenamefont {Lee}(1952)}]{Lee-Yang1}%
  \BibitemOpen
  \bibfield  {author} {\bibinfo {author} {\bibfnamefont {C.~N.}\ \bibnamefont
  {Yang}}\ and\ \bibinfo {author} {\bibfnamefont {T.~D.}\ \bibnamefont {Lee}},\
  }\bibfield  {title} {\enquote {\bibinfo {title} {Statistical theory of
  equations of state and phase transitions. i. theory of condensation},}\
  }\href {\doibase 10.1103/PhysRev.87.404} {\bibfield  {journal} {\bibinfo
  {journal} {Phys. Rev.}\ }\textbf {\bibinfo {volume} {87}},\ \bibinfo {pages}
  {404--409} (\bibinfo {year} {1952})}\BibitemShut {NoStop}%
\bibitem [{\citenamefont {Lee}\ and\ \citenamefont {Yang}(1952)}]{Lee-Yang2}%
  \BibitemOpen
  \bibfield  {author} {\bibinfo {author} {\bibfnamefont {T.~D.}\ \bibnamefont
  {Lee}}\ and\ \bibinfo {author} {\bibfnamefont {C.~N.}\ \bibnamefont {Yang}},\
  }\bibfield  {title} {\enquote {\bibinfo {title} {Statistical theory of
  equations of state and phase transitions. ii. lattice gas and ising model},}\
  }\href {\doibase 10.1103/PhysRev.87.410} {\bibfield  {journal} {\bibinfo
  {journal} {Phys. Rev.}\ }\textbf {\bibinfo {volume} {87}},\ \bibinfo {pages}
  {410--419} (\bibinfo {year} {1952})}\BibitemShut {NoStop}%
\bibitem [{\citenamefont {Utsumi}\ \emph {et~al.}(2013)\citenamefont {Utsumi},
  \citenamefont {Entin-Wohlman}, \citenamefont {Ueda},\ and\ \citenamefont
  {Aharony}}]{Utsumi_PRB}%
  \BibitemOpen
  \bibfield  {author} {\bibinfo {author} {\bibfnamefont {Y.}~\bibnamefont
  {Utsumi}}, \bibinfo {author} {\bibfnamefont {O.}~\bibnamefont
  {Entin-Wohlman}}, \bibinfo {author} {\bibfnamefont {A.}~\bibnamefont {Ueda}},
  \ and\ \bibinfo {author} {\bibfnamefont {A.}~\bibnamefont {Aharony}},\
  }\bibfield  {title} {\enquote {\bibinfo {title} {Full-counting statistics for
  molecular junctions: Fluctuation theorem and singularities},}\ }\href
  {\doibase 10.1103/PhysRevB.87.115407} {\bibfield  {journal} {\bibinfo
  {journal} {Phys. Rev. B}\ }\textbf {\bibinfo {volume} {87}},\ \bibinfo
  {pages} {115407} (\bibinfo {year} {2013})}\BibitemShut {NoStop}%
\bibitem [{\citenamefont {Peng}\ \emph {et~al.}(2015)\citenamefont {Peng},
  \citenamefont {Zhou}, \citenamefont {Wei}, \citenamefont {Cui}, \citenamefont
  {Du},\ and\ \citenamefont {Liu}}]{Peng_PRL}%
  \BibitemOpen
  \bibfield  {author} {\bibinfo {author} {\bibfnamefont {Xinhua}\ \bibnamefont
  {Peng}}, \bibinfo {author} {\bibfnamefont {Hui}\ \bibnamefont {Zhou}},
  \bibinfo {author} {\bibfnamefont {Bo-Bo}\ \bibnamefont {Wei}}, \bibinfo
  {author} {\bibfnamefont {Jiangyu}\ \bibnamefont {Cui}}, \bibinfo {author}
  {\bibfnamefont {Jiangfeng}\ \bibnamefont {Du}}, \ and\ \bibinfo {author}
  {\bibfnamefont {Ren-Bao}\ \bibnamefont {Liu}},\ }\bibfield  {title} {\enquote
  {\bibinfo {title} {Experimental observation of lee-yang zeros},}\ }\href
  {\doibase 10.1103/PhysRevLett.114.010601} {\bibfield  {journal} {\bibinfo
  {journal} {Phys. Rev. Lett.}\ }\textbf {\bibinfo {volume} {114}},\ \bibinfo
  {pages} {010601} (\bibinfo {year} {2015})}\BibitemShut {NoStop}%
\bibitem [{\citenamefont {Ivanov}\ and\ \citenamefont
  {Abanov}(2013)}]{Ivanov_PRE}%
  \BibitemOpen
  \bibfield  {author} {\bibinfo {author} {\bibfnamefont {Dmitri~A.}\
  \bibnamefont {Ivanov}}\ and\ \bibinfo {author} {\bibfnamefont {Alexander~G.}\
  \bibnamefont {Abanov}},\ }\bibfield  {title} {\enquote {\bibinfo {title}
  {Characterizing correlations with full counting statistics: Classical ising
  and quantum $xy$ spin chains},}\ }\href {\doibase 10.1103/PhysRevE.87.022114}
  {\bibfield  {journal} {\bibinfo  {journal} {Phys. Rev. E}\ }\textbf {\bibinfo
  {volume} {87}},\ \bibinfo {pages} {022114} (\bibinfo {year}
  {2013})}\BibitemShut {NoStop}%
\bibitem [{\citenamefont {Souto}\ \emph {et~al.}(2016)\citenamefont {Souto},
  \citenamefont {Mart\'{\i}n-Rodero},\ and\ \citenamefont
  {Yeyati}}]{SNS_Souto}%
  \BibitemOpen
  \bibfield  {author} {\bibinfo {author} {\bibfnamefont {R.~S.}\ \bibnamefont
  {Souto}}, \bibinfo {author} {\bibfnamefont {A.}~\bibnamefont
  {Mart\'{\i}n-Rodero}}, \ and\ \bibinfo {author} {\bibfnamefont {A.~L.}\
  \bibnamefont {Yeyati}},\ }\bibfield  {title} {\enquote {\bibinfo {title}
  {Andreev bound states formation and quasiparticle trapping in quench dynamics
  revealed by time-dependent counting statistics},}\ }\href {\doibase
  10.1103/PhysRevLett.117.267701} {\bibfield  {journal} {\bibinfo  {journal}
  {Phys. Rev. Lett.}\ }\textbf {\bibinfo {volume} {117}},\ \bibinfo {pages}
  {267701} (\bibinfo {year} {2016})}\BibitemShut {NoStop}%
\bibitem [{\citenamefont {Flindt}\ and\ \citenamefont
  {Garrahan}(2013)}]{Garrahan_PRL2013}%
  \BibitemOpen
  \bibfield  {author} {\bibinfo {author} {\bibfnamefont {Christian}\
  \bibnamefont {Flindt}}\ and\ \bibinfo {author} {\bibfnamefont {Juan~P.}\
  \bibnamefont {Garrahan}},\ }\bibfield  {title} {\enquote {\bibinfo {title}
  {Trajectory phase transitions, lee-yang zeros, and high-order cumulants in
  full counting statistics},}\ }\href {\doibase 10.1103/PhysRevLett.110.050601}
  {\bibfield  {journal} {\bibinfo  {journal} {Phys. Rev. Lett.}\ }\textbf
  {\bibinfo {volume} {110}},\ \bibinfo {pages} {050601} (\bibinfo {year}
  {2013})}\BibitemShut {NoStop}%
\bibitem [{\citenamefont {Hickey}\ \emph {et~al.}(2014)\citenamefont {Hickey},
  \citenamefont {Flindt},\ and\ \citenamefont {Garrahan}}]{Garrahan_PRE2014}%
  \BibitemOpen
  \bibfield  {author} {\bibinfo {author} {\bibfnamefont {James~M.}\
  \bibnamefont {Hickey}}, \bibinfo {author} {\bibfnamefont {Christian}\
  \bibnamefont {Flindt}}, \ and\ \bibinfo {author} {\bibfnamefont {Juan~P.}\
  \bibnamefont {Garrahan}},\ }\bibfield  {title} {\enquote {\bibinfo {title}
  {Intermittency and dynamical lee-yang zeros of open quantum systems},}\
  }\href {\doibase 10.1103/PhysRevE.90.062128} {\bibfield  {journal} {\bibinfo
  {journal} {Phys. Rev. E}\ }\textbf {\bibinfo {volume} {90}},\ \bibinfo
  {pages} {062128} (\bibinfo {year} {2014})}\BibitemShut {NoStop}%
\bibitem [{\citenamefont {Brandner}\ \emph {et~al.}(2017)\citenamefont
  {Brandner}, \citenamefont {Maisi}, \citenamefont {Pekola}, \citenamefont
  {Garrahan},\ and\ \citenamefont {Flindt}}]{Garrahan_PRL2017}%
  \BibitemOpen
  \bibfield  {author} {\bibinfo {author} {\bibfnamefont {Kay}\ \bibnamefont
  {Brandner}}, \bibinfo {author} {\bibfnamefont {Ville~F.}\ \bibnamefont
  {Maisi}}, \bibinfo {author} {\bibfnamefont {Jukka~P.}\ \bibnamefont
  {Pekola}}, \bibinfo {author} {\bibfnamefont {Juan~P.}\ \bibnamefont
  {Garrahan}}, \ and\ \bibinfo {author} {\bibfnamefont {Christian}\
  \bibnamefont {Flindt}},\ }\bibfield  {title} {\enquote {\bibinfo {title}
  {Experimental determination of dynamical lee-yang zeros},}\ }\href {\doibase
  10.1103/PhysRevLett.118.180601} {\bibfield  {journal} {\bibinfo  {journal}
  {Phys. Rev. Lett.}\ }\textbf {\bibinfo {volume} {118}},\ \bibinfo {pages}
  {180601} (\bibinfo {year} {2017})}\BibitemShut {NoStop}%
\bibitem [{\citenamefont {Flindt}\ \emph {et~al.}(2009)\citenamefont {Flindt},
  \citenamefont {Fricke}, \citenamefont {Hohls}, \citenamefont {Novotný},
  \citenamefont {Netočný}, \citenamefont {Brandes},\ and\ \citenamefont
  {Haug}}]{Flindt_PNAS}%
  \BibitemOpen
  \bibfield  {author} {\bibinfo {author} {\bibfnamefont {C.}~\bibnamefont
  {Flindt}}, \bibinfo {author} {\bibfnamefont {C.}~\bibnamefont {Fricke}},
  \bibinfo {author} {\bibfnamefont {F.}~\bibnamefont {Hohls}}, \bibinfo
  {author} {\bibfnamefont {T.}~\bibnamefont {Novotný}}, \bibinfo {author}
  {\bibfnamefont {K.}~\bibnamefont {Netočný}}, \bibinfo {author}
  {\bibfnamefont {T.}~\bibnamefont {Brandes}}, \ and\ \bibinfo {author}
  {\bibfnamefont {R.J.}\ \bibnamefont {Haug}},\ }\bibfield  {title} {\enquote
  {\bibinfo {title} {Universal oscillations in counting statistics},}\ }\href
  {\doibase 10.1073/pnas.0901002106} {\bibfield  {journal} {\bibinfo  {journal}
  {Proceedings of the National Academy of Sciences of the United States of
  America}\ }\textbf {\bibinfo {volume} {106}},\ \bibinfo {pages}
  {10116--10119} (\bibinfo {year} {2009})}\BibitemShut {NoStop}%
\bibitem [{\citenamefont {Levitov}(2002)}]{levitov}%
  \BibitemOpen
  \bibfield  {author} {\bibinfo {author} {\bibfnamefont {L.S.}\ \bibnamefont
  {Levitov}},\ }\href@noop {} {\emph {\bibinfo {title} {Quantum noise in
  Mesoscopic Physics}}}\ (\bibinfo  {publisher} {Kluwer Academic Press, New
  York},\ \bibinfo {year} {2002})\BibitemShut {NoStop}%
\bibitem [{\citenamefont {Cohen}\ \emph {et~al.}(2014)\citenamefont {Cohen},
  \citenamefont {Gull}, \citenamefont {Reichman},\ and\ \citenamefont
  {Millis}}]{CohenMillis_PRL2014}%
  \BibitemOpen
  \bibfield  {author} {\bibinfo {author} {\bibfnamefont {Guy}\ \bibnamefont
  {Cohen}}, \bibinfo {author} {\bibfnamefont {Emanuel}\ \bibnamefont {Gull}},
  \bibinfo {author} {\bibfnamefont {David~R.}\ \bibnamefont {Reichman}}, \ and\
  \bibinfo {author} {\bibfnamefont {Andrew~J.}\ \bibnamefont {Millis}},\
  }\bibfield  {title} {\enquote {\bibinfo {title} {Green's functions from
  real-time bold-line monte carlo calculations: Spectral properties of the
  nonequilibrium anderson impurity model},}\ }\href {\doibase
  10.1103/PhysRevLett.112.146802} {\bibfield  {journal} {\bibinfo  {journal}
  {Phys. Rev. Lett.}\ }\textbf {\bibinfo {volume} {112}},\ \bibinfo {pages}
  {146802} (\bibinfo {year} {2014})}\BibitemShut {NoStop}%
\bibitem [{\citenamefont {Chen}\ \emph {et~al.}(2016)\citenamefont {Chen},
  \citenamefont {Cohen}, \citenamefont {Millis},\ and\ \citenamefont
  {Reichman}}]{CohenMillis_PRB2016}%
  \BibitemOpen
  \bibfield  {author} {\bibinfo {author} {\bibfnamefont {Hsing-Ta}\
  \bibnamefont {Chen}}, \bibinfo {author} {\bibfnamefont {Guy}\ \bibnamefont
  {Cohen}}, \bibinfo {author} {\bibfnamefont {Andrew~J.}\ \bibnamefont
  {Millis}}, \ and\ \bibinfo {author} {\bibfnamefont {David~R.}\ \bibnamefont
  {Reichman}},\ }\bibfield  {title} {\enquote {\bibinfo {title}
  {Anderson-holstein model in two flavors of the noncrossing approximation},}\
  }\href {\doibase 10.1103/PhysRevB.93.174309} {\bibfield  {journal} {\bibinfo
  {journal} {Phys. Rev. B}\ }\textbf {\bibinfo {volume} {93}},\ \bibinfo
  {pages} {174309} (\bibinfo {year} {2016})}\BibitemShut {NoStop}%
\bibitem [{\citenamefont {Kamenev}(2011)}]{kamenev_book}%
  \BibitemOpen
  \bibfield  {author} {\bibinfo {author} {\bibfnamefont {A.}~\bibnamefont
  {Kamenev}},\ }\href {https://books.google.es/books?id=CwlrUepnla4C} {\emph
  {\bibinfo {title} {Field Theory of Non-Equilibrium Systems}}}\ (\bibinfo
  {publisher} {Cambridge University Press},\ \bibinfo {year}
  {2011})\BibitemShut {NoStop}%
\bibitem [{\citenamefont {Esposito}\ \emph {et~al.}(2009)\citenamefont
  {Esposito}, \citenamefont {Harbola},\ and\ \citenamefont
  {Mukamel}}]{mukamel}%
  \BibitemOpen
  \bibfield  {author} {\bibinfo {author} {\bibfnamefont {Massimiliano}\
  \bibnamefont {Esposito}}, \bibinfo {author} {\bibfnamefont {Upendra}\
  \bibnamefont {Harbola}}, \ and\ \bibinfo {author} {\bibfnamefont {Shaul}\
  \bibnamefont {Mukamel}},\ }\bibfield  {title} {\enquote {\bibinfo {title}
  {Nonequilibrium fluctuations, fluctuation theorems, and counting statistics
  in quantum systems},}\ }\href {\doibase 10.1103/RevModPhys.81.1665}
  {\bibfield  {journal} {\bibinfo  {journal} {Rev. Mod. Phys.}\ }\textbf
  {\bibinfo {volume} {81}},\ \bibinfo {pages} {1665--1702} (\bibinfo {year}
  {2009})}\BibitemShut {NoStop}%
\bibitem [{\citenamefont {Tang}\ \emph {et~al.}(2014)\citenamefont {Tang},
  \citenamefont {Xu},\ and\ \citenamefont {Wang}}]{Tang1}%
  \BibitemOpen
  \bibfield  {author} {\bibinfo {author} {\bibfnamefont {Gao-Min}\ \bibnamefont
  {Tang}}, \bibinfo {author} {\bibfnamefont {Fuming}\ \bibnamefont {Xu}}, \
  and\ \bibinfo {author} {\bibfnamefont {Jian}\ \bibnamefont {Wang}},\
  }\bibfield  {title} {\enquote {\bibinfo {title} {Waiting time distribution of
  quantum electronic transport in the transient regime},}\ }\href {\doibase
  10.1103/PhysRevB.89.205310} {\bibfield  {journal} {\bibinfo  {journal} {Phys.
  Rev. B}\ }\textbf {\bibinfo {volume} {89}},\ \bibinfo {pages} {205310}
  (\bibinfo {year} {2014})}\BibitemShut {NoStop}%
\bibitem [{\citenamefont {Tang}\ and\ \citenamefont {Wang}(2014)}]{Tang2}%
  \BibitemOpen
  \bibfield  {author} {\bibinfo {author} {\bibfnamefont {Gao-Min}\ \bibnamefont
  {Tang}}\ and\ \bibinfo {author} {\bibfnamefont {Jian}\ \bibnamefont {Wang}},\
  }\bibfield  {title} {\enquote {\bibinfo {title} {Full-counting statistics of
  charge and spin transport in the transient regime: A nonequilibrium green's
  function approach},}\ }\href {\doibase 10.1103/PhysRevB.90.195422} {\bibfield
   {journal} {\bibinfo  {journal} {Phys. Rev. B}\ }\textbf {\bibinfo {volume}
  {90}},\ \bibinfo {pages} {195422} (\bibinfo {year} {2014})}\BibitemShut
  {NoStop}%
\bibitem [{\citenamefont {Seoane~Souto}\ \emph {et~al.}(2015)\citenamefont
  {Seoane~Souto}, \citenamefont {Avriller}, \citenamefont {Monreal},
  \citenamefont {Mart\'{\i}n-Rodero},\ and\ \citenamefont
  {Levy~Yeyati}}]{Souto_PRB2015}%
  \BibitemOpen
  \bibfield  {author} {\bibinfo {author} {\bibfnamefont {R.}~\bibnamefont
  {Seoane~Souto}}, \bibinfo {author} {\bibfnamefont {R.}~\bibnamefont
  {Avriller}}, \bibinfo {author} {\bibfnamefont {R.~C.}\ \bibnamefont
  {Monreal}}, \bibinfo {author} {\bibfnamefont {A.}~\bibnamefont
  {Mart\'{\i}n-Rodero}}, \ and\ \bibinfo {author} {\bibfnamefont
  {A.}~\bibnamefont {Levy~Yeyati}},\ }\bibfield  {title} {\enquote {\bibinfo
  {title} {Transient dynamics and waiting time distribution of molecular
  junctions in the polaronic regime},}\ }\href {\doibase
  10.1103/PhysRevB.92.125435} {\bibfield  {journal} {\bibinfo  {journal} {Phys.
  Rev. B}\ }\textbf {\bibinfo {volume} {92}},\ \bibinfo {pages} {125435}
  (\bibinfo {year} {2015})}\BibitemShut {NoStop}%
\bibitem [{\citenamefont {Belzig}\ and\ \citenamefont
  {Nazarov}(2001)}]{belzig}%
  \BibitemOpen
  \bibfield  {author} {\bibinfo {author} {\bibfnamefont {W.}~\bibnamefont
  {Belzig}}\ and\ \bibinfo {author} {\bibfnamefont {Yu.~V.}\ \bibnamefont
  {Nazarov}},\ }\bibfield  {title} {\enquote {\bibinfo {title} {Full counting
  statistics of electron transfer between superconductors},}\ }\href {\doibase
  10.1103/PhysRevLett.87.197006} {\bibfield  {journal} {\bibinfo  {journal}
  {Phys. Rev. Lett.}\ }\textbf {\bibinfo {volume} {87}},\ \bibinfo {pages}
  {197006} (\bibinfo {year} {2001})}\BibitemShut {NoStop}%
\bibitem [{\citenamefont {Shelankov}\ and\ \citenamefont
  {Rammer}(2003)}]{ramer}%
  \BibitemOpen
  \bibfield  {author} {\bibinfo {author} {\bibfnamefont {A.}~\bibnamefont
  {Shelankov}}\ and\ \bibinfo {author} {\bibfnamefont {J.}~\bibnamefont
  {Rammer}},\ }\bibfield  {title} {\enquote {\bibinfo {title} {Charge transfer
  counting statistics revisited},}\ }\href
  {http://stacks.iop.org/0295-5075/63/i=4/a=485} {\bibfield  {journal}
  {\bibinfo  {journal} {EPL (Europhysics Letters)}\ }\textbf {\bibinfo {volume}
  {63}},\ \bibinfo {pages} {485} (\bibinfo {year} {2003})}\BibitemShut
  {NoStop}%
\bibitem [{\citenamefont {Hofer}\ and\ \citenamefont {Clerk}(2016)}]{clerk}%
  \BibitemOpen
  \bibfield  {author} {\bibinfo {author} {\bibfnamefont {Patrick~P.}\
  \bibnamefont {Hofer}}\ and\ \bibinfo {author} {\bibfnamefont {A.~A.}\
  \bibnamefont {Clerk}},\ }\bibfield  {title} {\enquote {\bibinfo {title}
  {Negative full counting statistics arise from interference effects},}\ }\href
  {\doibase 10.1103/PhysRevLett.116.013603} {\bibfield  {journal} {\bibinfo
  {journal} {Phys. Rev. Lett.}\ }\textbf {\bibinfo {volume} {116}},\ \bibinfo
  {pages} {013603} (\bibinfo {year} {2016})}\BibitemShut {NoStop}%
\bibitem [{\citenamefont {Hickey}\ \emph
  {et~al.}(2013{\natexlab{b}})\citenamefont {Hickey}, \citenamefont {Flindt},\
  and\ \citenamefont {Garrahan}}]{Garrahan_Flindt}%
  \BibitemOpen
  \bibfield  {author} {\bibinfo {author} {\bibfnamefont {James~M.}\
  \bibnamefont {Hickey}}, \bibinfo {author} {\bibfnamefont {Christian}\
  \bibnamefont {Flindt}}, \ and\ \bibinfo {author} {\bibfnamefont {Juan~P.}\
  \bibnamefont {Garrahan}},\ }\bibfield  {title} {\enquote {\bibinfo {title}
  {Trajectory phase transitions and dynamical lee-yang zeros of the
  glauber-ising chain},}\ }\href {\doibase 10.1103/PhysRevE.88.012119}
  {\bibfield  {journal} {\bibinfo  {journal} {Phys. Rev. E}\ }\textbf {\bibinfo
  {volume} {88}},\ \bibinfo {pages} {012119} (\bibinfo {year}
  {2013}{\natexlab{b}})}\BibitemShut {NoStop}%
\bibitem [{\citenamefont {Seoane~Souto}\ \emph {et~al.}(2017)\citenamefont
  {Seoane~Souto}, \citenamefont {Mart\'{\i}n-Rodero},\ and\ \citenamefont
  {Levy~Yeyati}}]{Souto_Fortschritte}%
  \BibitemOpen
  \bibfield  {author} {\bibinfo {author} {\bibfnamefont {R.}~\bibnamefont
  {Seoane~Souto}}, \bibinfo {author} {\bibfnamefont {A.}~\bibnamefont
  {Mart\'{\i}n-Rodero}}, \ and\ \bibinfo {author} {\bibfnamefont
  {A.}~\bibnamefont {Levy~Yeyati}},\ }\bibfield  {title} {\enquote {\bibinfo
  {title} {Analysis of universality in transient dynamics of coherent
  electronic transport},}\ }\href {\doibase 10.1002/prop.201600062} {\bibfield
  {journal} {\bibinfo  {journal} {Fortschr. Phys.}\ }\textbf {\bibinfo {volume}
  {65}},\ \bibinfo {pages} {1600062--n/a} (\bibinfo {year} {2017})}\BibitemShut
  {NoStop}%
\bibitem [{\citenamefont {Abramowitz}\ and\ \citenamefont
  {Stegun}(1964)}]{abramowitz_stegun}%
  \BibitemOpen
  \bibfield  {author} {\bibinfo {author} {\bibfnamefont {M.}~\bibnamefont
  {Abramowitz}}\ and\ \bibinfo {author} {\bibfnamefont {I.A.}\ \bibnamefont
  {Stegun}},\ }\href@noop {} {\emph {\bibinfo {title} {Handbook of Mathematical
  Functions: With Formulas, Graphs, and Mathematical Tables}}},\ Applied
  mathematics series\ (\bibinfo  {publisher} {Dover Publications},\ \bibinfo
  {year} {1964})\BibitemShut {NoStop}%
\bibitem [{\citenamefont {Berry}(2005)}]{Berry}%
  \BibitemOpen
  \bibfield  {author} {\bibinfo {author} {\bibfnamefont {M.V}\ \bibnamefont
  {Berry}},\ }\bibfield  {title} {\enquote {\bibinfo {title} {Universal
  oscillations of high derivatives},}\ }\href {\doibase 10.1098/rspa.2005.1446}
  {\bibfield  {journal} {\bibinfo  {journal} {Proc. R. Soc. of London A}\
  }\textbf {\bibinfo {volume} {461}},\ \bibinfo {pages} {1735--1751} (\bibinfo
  {year} {2005})}\BibitemShut {NoStop}%
\bibitem [{\citenamefont {Flindt}\ \emph {et~al.}(2010)\citenamefont {Flindt},
  \citenamefont {Novotn\'y}, \citenamefont {Braggio},\ and\ \citenamefont
  {Jauho}}]{Flindt_PRB2010}%
  \BibitemOpen
  \bibfield  {author} {\bibinfo {author} {\bibfnamefont {Christian}\
  \bibnamefont {Flindt}}, \bibinfo {author} {\bibfnamefont {Tom\'a\ifmmode
  \check{s}\else~\v{s}\fi{}}\ \bibnamefont {Novotn\'y}}, \bibinfo {author}
  {\bibfnamefont {Alessandro}\ \bibnamefont {Braggio}}, \ and\ \bibinfo
  {author} {\bibfnamefont {Antti-Pekka}\ \bibnamefont {Jauho}},\ }\bibfield
  {title} {\enquote {\bibinfo {title} {Counting statistics of transport through
  coulomb blockade nanostructures: High-order cumulants and non-markovian
  effects},}\ }\href {\doibase 10.1103/PhysRevB.82.155407} {\bibfield
  {journal} {\bibinfo  {journal} {Phys. Rev. B}\ }\textbf {\bibinfo {volume}
  {82}},\ \bibinfo {pages} {155407} (\bibinfo {year} {2010})}\BibitemShut
  {NoStop}%
\bibitem [{\citenamefont {Kambly}\ \emph {et~al.}(2011)\citenamefont {Kambly},
  \citenamefont {Flindt},\ and\ \citenamefont {B\"uttiker}}]{Kambly_factorial}%
  \BibitemOpen
  \bibfield  {author} {\bibinfo {author} {\bibfnamefont {Dania}\ \bibnamefont
  {Kambly}}, \bibinfo {author} {\bibfnamefont {Christian}\ \bibnamefont
  {Flindt}}, \ and\ \bibinfo {author} {\bibfnamefont {Markus}\ \bibnamefont
  {B\"uttiker}},\ }\bibfield  {title} {\enquote {\bibinfo {title} {Factorial
  cumulants reveal interactions in counting statistics},}\ }\href {\doibase
  10.1103/PhysRevB.83.075432} {\bibfield  {journal} {\bibinfo  {journal} {Phys.
  Rev. B}\ }\textbf {\bibinfo {volume} {83}},\ \bibinfo {pages} {075432}
  (\bibinfo {year} {2011})}\BibitemShut {NoStop}%
\bibitem [{\citenamefont {Stegmann}\ \emph {et~al.}(2015)\citenamefont
  {Stegmann}, \citenamefont {Sothmann}, \citenamefont {Hucht},\ and\
  \citenamefont {K\"onig}}]{Stegmann_factorial1}%
  \BibitemOpen
  \bibfield  {author} {\bibinfo {author} {\bibfnamefont {Philipp}\ \bibnamefont
  {Stegmann}}, \bibinfo {author} {\bibfnamefont {Bj\"orn}\ \bibnamefont
  {Sothmann}}, \bibinfo {author} {\bibfnamefont {Alfred}\ \bibnamefont
  {Hucht}}, \ and\ \bibinfo {author} {\bibfnamefont {J\"urgen}\ \bibnamefont
  {K\"onig}},\ }\bibfield  {title} {\enquote {\bibinfo {title} {Detection of
  interactions via generalized factorial cumulants in systems in and out of
  equilibrium},}\ }\href {\doibase 10.1103/PhysRevB.92.155413} {\bibfield
  {journal} {\bibinfo  {journal} {Phys. Rev. B}\ }\textbf {\bibinfo {volume}
  {92}},\ \bibinfo {pages} {155413} (\bibinfo {year} {2015})}\BibitemShut
  {NoStop}%
\bibitem [{\citenamefont {Stegmann}\ and\ \citenamefont
  {K\"onig}(2016)}]{Stegmann_factorial2}%
  \BibitemOpen
  \bibfield  {author} {\bibinfo {author} {\bibfnamefont {Philipp}\ \bibnamefont
  {Stegmann}}\ and\ \bibinfo {author} {\bibfnamefont {J\"urgen}\ \bibnamefont
  {K\"onig}},\ }\bibfield  {title} {\enquote {\bibinfo {title} {Short-time
  counting statistics of charge transfer in coulomb-blockade systems},}\ }\href
  {\doibase 10.1103/PhysRevB.94.125433} {\bibfield  {journal} {\bibinfo
  {journal} {Phys. Rev. B}\ }\textbf {\bibinfo {volume} {94}},\ \bibinfo
  {pages} {125433} (\bibinfo {year} {2016})}\BibitemShut {NoStop}%
\bibitem [{\citenamefont {Zazunov}\ \emph {et~al.}(2014)\citenamefont
  {Zazunov}, \citenamefont {Brunetti}, \citenamefont {Yeyati},\ and\
  \citenamefont {Egger}}]{Zazunov_PRB2014}%
  \BibitemOpen
  \bibfield  {author} {\bibinfo {author} {\bibfnamefont {A.}~\bibnamefont
  {Zazunov}}, \bibinfo {author} {\bibfnamefont {A.}~\bibnamefont {Brunetti}},
  \bibinfo {author} {\bibfnamefont {A.~L.}\ \bibnamefont {Yeyati}}, \ and\
  \bibinfo {author} {\bibfnamefont {R.}~\bibnamefont {Egger}},\ }\bibfield
  {title} {\enquote {\bibinfo {title} {Quasiparticle trapping, andreev level
  population dynamics, and charge imbalance in superconducting weak links},}\
  }\href {\doibase 10.1103/PhysRevB.90.104508} {\bibfield  {journal} {\bibinfo
  {journal} {Phys. Rev. B}\ }\textbf {\bibinfo {volume} {90}},\ \bibinfo
  {pages} {104508} (\bibinfo {year} {2014})}\BibitemShut {NoStop}%
\bibitem [{\citenamefont {Abanov}\ and\ \citenamefont
  {Ivanov}(2008)}]{Abanov_PRL2008}%
  \BibitemOpen
  \bibfield  {author} {\bibinfo {author} {\bibfnamefont {A.~G.}\ \bibnamefont
  {Abanov}}\ and\ \bibinfo {author} {\bibfnamefont {D.~A.}\ \bibnamefont
  {Ivanov}},\ }\bibfield  {title} {\enquote {\bibinfo {title} {Allowed charge
  transfers between coherent conductors driven by a time-dependent
  scatterer},}\ }\href {\doibase 10.1103/PhysRevLett.100.086602} {\bibfield
  {journal} {\bibinfo  {journal} {Phys. Rev. Lett.}\ }\textbf {\bibinfo
  {volume} {100}},\ \bibinfo {pages} {086602} (\bibinfo {year}
  {2008})}\BibitemShut {NoStop}%
\bibitem [{\citenamefont {Ivanov}\ and\ \citenamefont
  {Abanov}(2010)}]{Ivanov_europysics}%
  \BibitemOpen
  \bibfield  {author} {\bibinfo {author} {\bibfnamefont {D.~A.}\ \bibnamefont
  {Ivanov}}\ and\ \bibinfo {author} {\bibfnamefont {A.~G.}\ \bibnamefont
  {Abanov}},\ }\bibfield  {title} {\enquote {\bibinfo {title} {Phase
  transitions in full counting statistics for periodic pumping},}\ }\href
  {http://stacks.iop.org/0295-5075/92/i=3/a=37008} {\bibfield  {journal}
  {\bibinfo  {journal} {EPL (Europhysics Letters)}\ }\textbf {\bibinfo {volume}
  {92}},\ \bibinfo {pages} {37008} (\bibinfo {year} {2010})}\BibitemShut
  {NoStop}%
\bibitem [{\citenamefont {Perfetto}\ \emph {et~al.}(2009)\citenamefont
  {Perfetto}, \citenamefont {Stefanucci},\ and\ \citenamefont
  {Cini}}]{Stefanucci_PRB_2009}%
  \BibitemOpen
  \bibfield  {author} {\bibinfo {author} {\bibfnamefont {Enrico}\ \bibnamefont
  {Perfetto}}, \bibinfo {author} {\bibfnamefont {Gianluca}\ \bibnamefont
  {Stefanucci}}, \ and\ \bibinfo {author} {\bibfnamefont {Michele}\
  \bibnamefont {Cini}},\ }\bibfield  {title} {\enquote {\bibinfo {title}
  {Equilibrium and time-dependent josephson current in one-dimensional
  superconducting junctions},}\ }\href {\doibase 10.1103/PhysRevB.80.205408}
  {\bibfield  {journal} {\bibinfo  {journal} {Phys. Rev. B}\ }\textbf {\bibinfo
  {volume} {80}},\ \bibinfo {pages} {205408} (\bibinfo {year}
  {2009})}\BibitemShut {NoStop}%
\bibitem [{\citenamefont {Stefanucci}\ \emph {et~al.}(2010)\citenamefont
  {Stefanucci}, \citenamefont {Perfetto},\ and\ \citenamefont
  {Cini}}]{Stefanucci_PRB_2010}%
  \BibitemOpen
  \bibfield  {author} {\bibinfo {author} {\bibfnamefont {Gianluca}\
  \bibnamefont {Stefanucci}}, \bibinfo {author} {\bibfnamefont {Enrico}\
  \bibnamefont {Perfetto}}, \ and\ \bibinfo {author} {\bibfnamefont {Michele}\
  \bibnamefont {Cini}},\ }\bibfield  {title} {\enquote {\bibinfo {title}
  {Time-dependent quantum transport with superconducting leads: A
  discrete-basis kohn-sham formulation and propagation scheme},}\ }\href
  {\doibase 10.1103/PhysRevB.81.115446} {\bibfield  {journal} {\bibinfo
  {journal} {Phys. Rev. B}\ }\textbf {\bibinfo {volume} {81}},\ \bibinfo
  {pages} {115446} (\bibinfo {year} {2010})}\BibitemShut {NoStop}%
\bibitem [{\citenamefont {Albrecht}\ \emph {et~al.}(2013)\citenamefont
  {Albrecht}, \citenamefont {Soller}, \citenamefont {Mühlbacher},\ and\
  \citenamefont {Komnik}}]{Albrecht_2013}%
  \BibitemOpen
  \bibfield  {author} {\bibinfo {author} {\bibfnamefont {K.F.}\ \bibnamefont
  {Albrecht}}, \bibinfo {author} {\bibfnamefont {H.}~\bibnamefont {Soller}},
  \bibinfo {author} {\bibfnamefont {L.}~\bibnamefont {Mühlbacher}}, \ and\
  \bibinfo {author} {\bibfnamefont {A.}~\bibnamefont {Komnik}},\ }\bibfield
  {title} {\enquote {\bibinfo {title} {Transient dynamics and steady state
  behavior of the anderson–holstein model with a superconducting lead},}\
  }\href {\doibase 10.1016/j.physe.2013.05.019} {\bibfield  {journal} {\bibinfo
   {journal} {Phys. E}\ }\textbf {\bibinfo {volume} {54}},\ \bibinfo {pages}
  {15--23} (\bibinfo {year} {2013})}\BibitemShut {NoStop}%
\bibitem [{\citenamefont {Weston}\ and\ \citenamefont
  {Waintal}(2016)}]{Weston_PRB_2016}%
  \BibitemOpen
  \bibfield  {author} {\bibinfo {author} {\bibfnamefont {Joseph}\ \bibnamefont
  {Weston}}\ and\ \bibinfo {author} {\bibfnamefont {Xavier}\ \bibnamefont
  {Waintal}},\ }\bibfield  {title} {\enquote {\bibinfo {title} {Linear-scaling
  source-sink algorithm for simulating time-resolved quantum transport and
  superconductivity},}\ }\href {\doibase 10.1103/PhysRevB.93.134506} {\bibfield
   {journal} {\bibinfo  {journal} {Phys. Rev. B}\ }\textbf {\bibinfo {volume}
  {93}},\ \bibinfo {pages} {134506} (\bibinfo {year} {2016})}\BibitemShut
  {NoStop}%
\bibitem [{\citenamefont {Taranko}\ and\ \citenamefont
  {Domanski}(2017)}]{Taranko_arXiv_2017}%
  \BibitemOpen
  \bibfield  {author} {\bibinfo {author} {\bibfnamefont {R.}~\bibnamefont
  {Taranko}}\ and\ \bibinfo {author} {\bibfnamefont {T.}~\bibnamefont
  {Domanski}},\ }\bibfield  {title} {\enquote {\bibinfo {title} {How long does
  it take to form the andreev quasiparticles?}}\ }\href
  {https://arxiv.org/abs/1705.08755} {\bibfield  {journal} {\bibinfo  {journal}
  {arXiv:1705.08755}\ } (\bibinfo {year} {2017})}\BibitemShut {NoStop}%
\bibitem [{\citenamefont {Bratus}\ \emph {et~al.}(1995)\citenamefont {Bratus},
  \citenamefont {Shumeiko},\ and\ \citenamefont {Wendin}}]{Bratus_PRL1995}%
  \BibitemOpen
  \bibfield  {author} {\bibinfo {author} {\bibfnamefont {E.~N.}\ \bibnamefont
  {Bratus}}, \bibinfo {author} {\bibfnamefont {V.~S.}\ \bibnamefont
  {Shumeiko}}, \ and\ \bibinfo {author} {\bibfnamefont {G.}~\bibnamefont
  {Wendin}},\ }\bibfield  {title} {\enquote {\bibinfo {title} {Theory of
  subharmonic gap structure in superconducting mesoscopic tunnel contacts},}\
  }\href {\doibase 10.1103/PhysRevLett.74.2110} {\bibfield  {journal} {\bibinfo
   {journal} {Phys. Rev. Lett.}\ }\textbf {\bibinfo {volume} {74}},\ \bibinfo
  {pages} {2110--2113} (\bibinfo {year} {1995})}\BibitemShut {NoStop}%
\bibitem [{\citenamefont {Averin}\ and\ \citenamefont
  {Bardas}(1995)}]{Averin_PRL_1995}%
  \BibitemOpen
  \bibfield  {author} {\bibinfo {author} {\bibfnamefont {D.}~\bibnamefont
  {Averin}}\ and\ \bibinfo {author} {\bibfnamefont {A.}~\bibnamefont
  {Bardas}},\ }\bibfield  {title} {\enquote {\bibinfo {title} {ac josephson
  effect in a single quantum channel},}\ }\href {\doibase
  10.1103/PhysRevLett.75.1831} {\bibfield  {journal} {\bibinfo  {journal}
  {Phys. Rev. Lett.}\ }\textbf {\bibinfo {volume} {75}},\ \bibinfo {pages}
  {1831--1834} (\bibinfo {year} {1995})}\BibitemShut {NoStop}%
\bibitem [{\citenamefont {Cuevas}\ \emph {et~al.}(1996)\citenamefont {Cuevas},
  \citenamefont {Mart\'{\i}n-Rodero},\ and\ \citenamefont
  {Yeyati}}]{Cuevas_PRB_1996}%
  \BibitemOpen
  \bibfield  {author} {\bibinfo {author} {\bibfnamefont {J.~C.}\ \bibnamefont
  {Cuevas}}, \bibinfo {author} {\bibfnamefont {A.}~\bibnamefont
  {Mart\'{\i}n-Rodero}}, \ and\ \bibinfo {author} {\bibfnamefont {A.~Levy}\
  \bibnamefont {Yeyati}},\ }\bibfield  {title} {\enquote {\bibinfo {title}
  {Hamiltonian approach to the transport properties of superconducting quantum
  point contacts},}\ }\href {\doibase 10.1103/PhysRevB.54.7366} {\bibfield
  {journal} {\bibinfo  {journal} {Phys. Rev. B}\ }\textbf {\bibinfo {volume}
  {54}},\ \bibinfo {pages} {7366--7379} (\bibinfo {year} {1996})}\BibitemShut
  {NoStop}%
\bibitem [{\citenamefont {Yeyati}\ \emph {et~al.}(1997)\citenamefont {Yeyati},
  \citenamefont {Cuevas}, \citenamefont {L\'opez-D\'avalos},\ and\
  \citenamefont {Mart\'{\i}n-Rodero}}]{Yeyati_PRB_1997}%
  \BibitemOpen
  \bibfield  {author} {\bibinfo {author} {\bibfnamefont {A.~Levy}\ \bibnamefont
  {Yeyati}}, \bibinfo {author} {\bibfnamefont {J.~C.}\ \bibnamefont {Cuevas}},
  \bibinfo {author} {\bibfnamefont {A.}~\bibnamefont {L\'opez-D\'avalos}}, \
  and\ \bibinfo {author} {\bibfnamefont {A.}~\bibnamefont
  {Mart\'{\i}n-Rodero}},\ }\bibfield  {title} {\enquote {\bibinfo {title}
  {Resonant tunneling through a small quantum dot coupled to superconducting
  leads},}\ }\href {\doibase 10.1103/PhysRevB.55.R6137} {\bibfield  {journal}
  {\bibinfo  {journal} {Phys. Rev. B}\ }\textbf {\bibinfo {volume} {55}},\
  \bibinfo {pages} {R6137--R6140} (\bibinfo {year} {1997})}\BibitemShut
  {NoStop}%
\bibitem [{\citenamefont {Johansson}\ \emph {et~al.}(1999)\citenamefont
  {Johansson}, \citenamefont {Bratus}, \citenamefont {Shumeiko},\ and\
  \citenamefont {Wendin}}]{Johansson_PRB1999}%
  \BibitemOpen
  \bibfield  {author} {\bibinfo {author} {\bibfnamefont {G.}~\bibnamefont
  {Johansson}}, \bibinfo {author} {\bibfnamefont {E.~N.}\ \bibnamefont
  {Bratus}}, \bibinfo {author} {\bibfnamefont {V.~S.}\ \bibnamefont
  {Shumeiko}}, \ and\ \bibinfo {author} {\bibfnamefont {G.}~\bibnamefont
  {Wendin}},\ }\bibfield  {title} {\enquote {\bibinfo {title} {Resonant
  multiple andreev reflections in mesoscopic superconducting junctions},}\
  }\href {\doibase 10.1103/PhysRevB.60.1382} {\bibfield  {journal} {\bibinfo
  {journal} {Phys. Rev. B}\ }\textbf {\bibinfo {volume} {60}},\ \bibinfo
  {pages} {1382--1393} (\bibinfo {year} {1999})}\BibitemShut {NoStop}%
\bibitem [{\citenamefont {Yeyati}\ \emph {et~al.}(2003)\citenamefont {Yeyati},
  \citenamefont {Mart\'{\i}n-Rodero},\ and\ \citenamefont
  {Vecino}}]{Yeyati_PRL_2003}%
  \BibitemOpen
  \bibfield  {author} {\bibinfo {author} {\bibfnamefont {A.~Levy}\ \bibnamefont
  {Yeyati}}, \bibinfo {author} {\bibfnamefont {A.}~\bibnamefont
  {Mart\'{\i}n-Rodero}}, \ and\ \bibinfo {author} {\bibfnamefont
  {E.}~\bibnamefont {Vecino}},\ }\bibfield  {title} {\enquote {\bibinfo {title}
  {Nonequilibrium dynamics of andreev states in the kondo regime},}\ }\href
  {\doibase 10.1103/PhysRevLett.91.266802} {\bibfield  {journal} {\bibinfo
  {journal} {Phys. Rev. Lett.}\ }\textbf {\bibinfo {volume} {91}},\ \bibinfo
  {pages} {266802} (\bibinfo {year} {2003})}\BibitemShut {NoStop}%
\bibitem [{\citenamefont {Cuevas}\ \emph {et~al.}(1999)\citenamefont {Cuevas},
  \citenamefont {Mart\'{\i}n-Rodero},\ and\ \citenamefont
  {Yeyati}}]{Cuevas_PRL_1999}%
  \BibitemOpen
  \bibfield  {author} {\bibinfo {author} {\bibfnamefont {J.~C.}\ \bibnamefont
  {Cuevas}}, \bibinfo {author} {\bibfnamefont {A.}~\bibnamefont
  {Mart\'{\i}n-Rodero}}, \ and\ \bibinfo {author} {\bibfnamefont {A.~Levy}\
  \bibnamefont {Yeyati}},\ }\bibfield  {title} {\enquote {\bibinfo {title}
  {Shot noise and coherent multiple charge transfer in superconducting quantum
  point contacts},}\ }\href {\doibase 10.1103/PhysRevLett.82.4086} {\bibfield
  {journal} {\bibinfo  {journal} {Phys. Rev. Lett.}\ }\textbf {\bibinfo
  {volume} {82}},\ \bibinfo {pages} {4086--4089} (\bibinfo {year}
  {1999})}\BibitemShut {NoStop}%
\bibitem [{\citenamefont {Cuevas}\ and\ \citenamefont
  {Belzig}(2003)}]{Cuevas_PRL_2003}%
  \BibitemOpen
  \bibfield  {author} {\bibinfo {author} {\bibfnamefont {J.~C.}\ \bibnamefont
  {Cuevas}}\ and\ \bibinfo {author} {\bibfnamefont {W.}~\bibnamefont
  {Belzig}},\ }\bibfield  {title} {\enquote {\bibinfo {title} {Full counting
  statistics of multiple andreev reflections},}\ }\href {\doibase
  10.1103/PhysRevLett.91.187001} {\bibfield  {journal} {\bibinfo  {journal}
  {Phys. Rev. Lett.}\ }\textbf {\bibinfo {volume} {91}},\ \bibinfo {pages}
  {187001} (\bibinfo {year} {2003})}\BibitemShut {NoStop}%
\bibitem [{\citenamefont {Cuevas}\ and\ \citenamefont
  {Belzig}(2004)}]{Cuevas_PRB_2004}%
  \BibitemOpen
  \bibfield  {author} {\bibinfo {author} {\bibfnamefont {J.~C.}\ \bibnamefont
  {Cuevas}}\ and\ \bibinfo {author} {\bibfnamefont {W.}~\bibnamefont
  {Belzig}},\ }\bibfield  {title} {\enquote {\bibinfo {title} {dc transport in
  superconducting point contacts: A full-counting-statistics view},}\ }\href
  {\doibase 10.1103/PhysRevB.70.214512} {\bibfield  {journal} {\bibinfo
  {journal} {Phys. Rev. B}\ }\textbf {\bibinfo {volume} {70}},\ \bibinfo
  {pages} {214512} (\bibinfo {year} {2004})}\BibitemShut {NoStop}%
\bibitem [{\citenamefont {Sch\"onenberger}()}]{privateCom}%
  \BibitemOpen
  \bibfield  {author} {\bibinfo {author} {\bibfnamefont {Christian}\
  \bibnamefont {Sch\"onenberger}},\ }\href@noop {} {\bibinfo  {journal}
  {Private communication}\ }\BibitemShut {NoStop}%
\bibitem [{\citenamefont {Freyn}\ \emph {et~al.}(2011)\citenamefont {Freyn},
  \citenamefont {Dou\ifmmode~\mbox{\c{c}}\else \c{c}\fi{}ot}, \citenamefont
  {Feinberg},\ and\ \citenamefont {M\'elin}}]{Freyn_PRL_2011}%
  \BibitemOpen
\bibfield  {journal} {  }\bibfield  {author} {\bibinfo {author} {\bibfnamefont
  {Axel}\ \bibnamefont {Freyn}}, \bibinfo {author} {\bibfnamefont {Benoit}\
  \bibnamefont {Dou\ifmmode~\mbox{\c{c}}\else \c{c}\fi{}ot}}, \bibinfo {author}
  {\bibfnamefont {Denis}\ \bibnamefont {Feinberg}}, \ and\ \bibinfo {author}
  {\bibfnamefont {R\'egis}\ \bibnamefont {M\'elin}},\ }\bibfield  {title}
  {\enquote {\bibinfo {title} {Production of nonlocal quartets and
  phase-sensitive entanglement in a superconducting beam splitter},}\ }\href
  {\doibase 10.1103/PhysRevLett.106.257005} {\bibfield  {journal} {\bibinfo
  {journal} {Phys. Rev. Lett.}\ }\textbf {\bibinfo {volume} {106}},\ \bibinfo
  {pages} {257005} (\bibinfo {year} {2011})}\BibitemShut {NoStop}%
\bibitem [{\citenamefont {Jonckheere}\ \emph {et~al.}(2013)\citenamefont
  {Jonckheere}, \citenamefont {Rech}, \citenamefont {Martin}, \citenamefont
  {Dou\mbox{\c{c}}ot}, \citenamefont {Feinberg},\ and\ \citenamefont
  {M\'elin}}]{Melin_PRB_2013}%
  \BibitemOpen
  \bibfield  {author} {\bibinfo {author} {\bibfnamefont {T.}~\bibnamefont
  {Jonckheere}}, \bibinfo {author} {\bibfnamefont {J.}~\bibnamefont {Rech}},
  \bibinfo {author} {\bibfnamefont {T.}~\bibnamefont {Martin}}, \bibinfo
  {author} {\bibfnamefont {B.}~\bibnamefont {Dou\mbox{\c{c}}ot}}, \bibinfo
  {author} {\bibfnamefont {D.}~\bibnamefont {Feinberg}}, \ and\ \bibinfo
  {author} {\bibfnamefont {R.}~\bibnamefont {M\'elin}},\ }\bibfield  {title}
  {\enquote {\bibinfo {title} {Multipair dc josephson resonances in a biased
  all-superconducting bijunction},}\ }\href {\doibase
  10.1103/PhysRevB.87.214501} {\bibfield  {journal} {\bibinfo  {journal} {Phys.
  Rev. B}\ }\textbf {\bibinfo {volume} {87}},\ \bibinfo {pages} {214501}
  (\bibinfo {year} {2013})}\BibitemShut {NoStop}%
\bibitem [{\citenamefont {Pfeffer}\ \emph {et~al.}(2014)\citenamefont
  {Pfeffer}, \citenamefont {Duvauchelle}, \citenamefont {Courtois},
  \citenamefont {M\'elin}, \citenamefont {Feinberg},\ and\ \citenamefont
  {Lefloch}}]{Pfeffer_PRB_2014}%
  \BibitemOpen
  \bibfield  {author} {\bibinfo {author} {\bibfnamefont {A.~H.}\ \bibnamefont
  {Pfeffer}}, \bibinfo {author} {\bibfnamefont {J.~E.}\ \bibnamefont
  {Duvauchelle}}, \bibinfo {author} {\bibfnamefont {H.}~\bibnamefont
  {Courtois}}, \bibinfo {author} {\bibfnamefont {R.}~\bibnamefont {M\'elin}},
  \bibinfo {author} {\bibfnamefont {D.}~\bibnamefont {Feinberg}}, \ and\
  \bibinfo {author} {\bibfnamefont {F.}~\bibnamefont {Lefloch}},\ }\bibfield
  {title} {\enquote {\bibinfo {title} {Subgap structure in the conductance of a
  three-terminal josephson junction},}\ }\href {\doibase
  10.1103/PhysRevB.90.075401} {\bibfield  {journal} {\bibinfo  {journal} {Phys.
  Rev. B}\ }\textbf {\bibinfo {volume} {90}},\ \bibinfo {pages} {075401}
  (\bibinfo {year} {2014})}\BibitemShut {NoStop}%
\bibitem [{\citenamefont {Nandkishore}\ and\ \citenamefont
  {Huse}(2015)}]{Nandkishore}%
  \BibitemOpen
  \bibfield  {author} {\bibinfo {author} {\bibfnamefont {R}~\bibnamefont
  {Nandkishore}}\ and\ \bibinfo {author} {\bibfnamefont {D~A}\ \bibnamefont
  {Huse}},\ }\bibfield  {title} {\enquote {\bibinfo {title} {Nonequilibrium
  fluctuations, fluctuation theorems, and counting statistics in quantum
  systems},}\ }\href
  {http://dx.doi.org/10.1146/annurev-conmatphys-031214-014548} {\bibfield
  {journal} {\bibinfo  {journal} {Ann. Rev. Condens. Matter Phys.}\ }\textbf
  {\bibinfo {volume} {6}},\ \bibinfo {pages} {201} (\bibinfo {year}
  {2015})}\BibitemShut {NoStop}%
\bibitem [{\citenamefont {Vasseur}\ and\ \citenamefont
  {Moore}(2016)}]{Vaseur_JSM}%
  \BibitemOpen
  \bibfield  {author} {\bibinfo {author} {\bibfnamefont {Romain}\ \bibnamefont
  {Vasseur}}\ and\ \bibinfo {author} {\bibfnamefont {Joel~E}\ \bibnamefont
  {Moore}},\ }\bibfield  {title} {\enquote {\bibinfo {title} {Nonequilibrium
  quantum dynamics and transport: from integrability to many-body
  localization},}\ }\href {http://stacks.iop.org/1742-5468/2016/i=6/a=064010}
  {\bibfield  {journal} {\bibinfo  {journal} {Journal of Statistical Mechanics:
  Theory and Experiment}\ }\textbf {\bibinfo {volume} {2016}},\ \bibinfo
  {pages} {064010} (\bibinfo {year} {2016})}\BibitemShut {NoStop}%
\bibitem [{\citenamefont {Jauho}\ \emph {et~al.}(1994)\citenamefont {Jauho},
  \citenamefont {Wingreen},\ and\ \citenamefont {Meir}}]{Wingreen_PRB1994}%
  \BibitemOpen
  \bibfield  {author} {\bibinfo {author} {\bibfnamefont {Antti-Pekka}\
  \bibnamefont {Jauho}}, \bibinfo {author} {\bibfnamefont {Ned~S.}\
  \bibnamefont {Wingreen}}, \ and\ \bibinfo {author} {\bibfnamefont {Yigal}\
  \bibnamefont {Meir}},\ }\bibfield  {title} {\enquote {\bibinfo {title}
  {Time-dependent transport in interacting and noninteracting
  resonant-tunneling systems},}\ }\href {\doibase 10.1103/PhysRevB.50.5528}
  {\bibfield  {journal} {\bibinfo  {journal} {Phys. Rev. B}\ }\textbf {\bibinfo
  {volume} {50}},\ \bibinfo {pages} {5528--5544} (\bibinfo {year}
  {1994})}\BibitemShut {NoStop}%
\bibitem [{\citenamefont {Mart\'in-Rodero}\ and\ \citenamefont
  {Yeyati}(2011)}]{Martin_Advances}%
  \BibitemOpen
  \bibfield  {author} {\bibinfo {author} {\bibfnamefont {A.}~\bibnamefont
  {Mart\'in-Rodero}}\ and\ \bibinfo {author} {\bibfnamefont {A.~Levy}\
  \bibnamefont {Yeyati}},\ }\bibfield  {title} {\enquote {\bibinfo {title}
  {Josephson and andreev transport through quantum dots},}\ }\href {\doibase
  10.1080/00018732.2011.624266} {\bibfield  {journal} {\bibinfo  {journal}
  {Advances in Physics}\ }\textbf {\bibinfo {volume} {60}},\ \bibinfo {pages}
  {899--958} (\bibinfo {year} {2011})}\BibitemShut {NoStop}%
\bibitem [{\citenamefont {Gardiner}(2009)}]{Gardiner_book}%
  \BibitemOpen
  \bibfield  {author} {\bibinfo {author} {\bibfnamefont {C.}~\bibnamefont
  {Gardiner}},\ }\href@noop {} {\emph {\bibinfo {title} {Stochastic Methods: A
  Handbook for the Natural and Social Sciences}}},\ Springer Series in
  Synergetics\ (\bibinfo  {publisher} {Springer Berlin Heidelberg},\ \bibinfo
  {year} {2009})\BibitemShut {NoStop}%
\end{thebibliography}%
\appendix
\section{Analytical results for the mean charge in the central region}
\label{sec::SPA}
For the calculation of the mean charge and current it is convenient to use the Keldysh formalism in the triangular form, where only the
retarded, advanced and Keldysh $+-$ components are involved in the Dyson
equation. For an abrupt connection between the leads and the central region, the retarded (advanced) Green functions have
a simple form in Nambu space
\begin{equation}
\hat{G}^{R(A)}(t,t') = \theta(t)\theta(t') \hat{G}_{stat}^{R(A)}(t-t')\,,
\end{equation}
where $\hat{G}_{stat}^{R(A)}(t-t')$ denotes the stationary retarded (advanced) Green function. This expression is completely general for an abrupt connection in the absence of interactions, and 
it reduces to the expression provided in Ref. \cite{Wingreen_PRB1994} for normal electrodes. In the frequency domain the stationary retarded (advanced) Green function can be written as 
\begin{widetext}
\begin{equation}
\hat{G}_{stat}^{R(A)}(\omega) = \left(\begin{array}{cc} \omega - \epsilon_0 - \Gamma g_{11}^{R(A)} & (\Gamma_L e^{i\phi_L}+\Gamma_R e^{i\phi_R}) g_{12}^{R(A)} \\
 (\Gamma_L e^{-i\phi_L}+\Gamma_R e^{-i\phi_R}) g_{12}^{R(A)} & \omega + \epsilon_0 - \Gamma g_{22}^{R(A)}\end{array}\right)^{-1}
\end{equation}
\end{widetext}

where $g^{R(A)}_{\alpha\beta}$ are the BCS Green functions of the uncoupled electrodes \cite{Martin_Advances}.

$\hat{G}_{stat}^{R(A)}(\omega)$ have poles for $|\omega|\leq\Delta$ plane, which correspond to the ABSs at
$\pm \epsilon_A(\phi)$. In the limit $\Gamma/\Delta \ll 1$ the contribution from the
continuum spectrum for $|\omega|>\Delta$ becomes negligible and $\hat{G}^{R,A}(t,t')$ can be approximated by

\begin{equation}
\hat{G}^{R(A)}(t,t') \simeq \theta(t)\theta(t') \sum_{\pm} \left(\begin{array}{cc}
p_{\pm} & \pm p_{12} \\
\pm p^*_{12} & p_{\pm} \end{array}\right)e^{\pm i\epsilon_A(t-t')}
\label{retarded}
\end{equation}
where $|p_{12}|=\sqrt{p_+p_-}$. The ABS energies and the weights $p_{\pm}$ adopt a simple form when
$\epsilon_0 < \Gamma$, i.e. $\epsilon_A = \sqrt{\epsilon_0^2 + \Gamma^2 \cos^2(\phi/2) + (\Gamma_L-\Gamma_R)^2\sin^2(\phi/2)}$
and 
\begin{equation}
 P_\pm=\frac{\Delta}{2\sqrt{\Delta^2+\epsilon_{A}^2}}\left(1-\frac{\Gamma}{\Delta}\mp\frac{\epsilon_0}{\epsilon_A}\right)\,.
\end{equation}

The time-dependent level charge can then be obtained through the Dyson
equation for the Keldysh Green function $\hat{G}^{+-}$,

\begin{equation}
\hat{G}^{+-}= \hat{G}^R\hat{\Sigma}^{+-}
\hat{G}^A+\left(1+\hat{G}^R\hat{\Sigma}^R\right)\hat{g}_{0}^{+-}\left(1+\hat{\Sigma}^A\hat{G}^A\right)\,,
\label{keldysh_SPA} 
\end{equation}
where $\hat{g}_{0}^{+-}$ is the Keldysh Green function of the uncoupled central level in Nambu space,
\begin{equation}
 \hat{g}_{0}^{+-}(t,t') = ie^{-i\epsilon_0(t-t')}\left(\begin{array}{cc} n_\uparrow(0) & 0 \\
 0 & 1-n_\downarrow(0)\end{array}\right)\,.
\end{equation}
$\hat{\Sigma}^{+-}$ is the self energy coupling the level to the electrodes,
\begin{equation}
 \hat{\Sigma}^{+-}=\sum_{\nu}\left(G^R\Sigma^R\right)_{0\nu}g_{\nu}^{+-}\left(\Sigma^A G^A\right)_{\nu0}\,,
\end{equation}
where $g_{\nu}^{+-}(t,t')$ are the BCS Green functions of the superconducting electrodes, given in the supplemental material of Ref. \cite{SNS_Souto}.

For simplicity, the time arguments in Eq. (\ref{keldysh_SPA}) have not
been included and all the products represent time convolutions. For an initial condition $(n_\uparrow(0),n_\downarrow(0))=(0,1)$ only the first term in the equation contributes to $n_\uparrow(t)$. 
Substitution of Eq. \ref{retarded} in Eq. \ref{keldysh_SPA} then yields

\begin{widetext}
\begin{eqnarray}
n_{\uparrow}(t) &= & \sum_{\pm} n_{\pm}(t) + n_{+-} \nonumber \\
n_{\pm}(t) &=& -\frac{2 p_{\pm}}{\pi}\int_{-\infty}^{-\Delta} \frac{\omega\Gamma\mp 4p_+p_- \Delta\cdot\epsilon_A}{\sqrt{\omega^2-\Delta^2} \omega_{\mp}}\left[1-\cos\left(\omega_{\mp}t\right)\right]d\omega \nonumber\\
n_{+-} &=&  \frac{2(p_+-p_-)}{\pi}\int_{-\infty}^{-\Delta} \frac{4p_+p_-\Delta\cdot\epsilon_A}{\omega_+\omega_-\sqrt{\omega^2-\Delta^2}}\left[1 + 
\cos\left(2\epsilon_At\right)-\cos(\omega_+t) - \cos(\omega_-t)\right] d\omega
\end{eqnarray}
where $\omega_{\pm} = \omega\pm\epsilon_A$ (the spin down population is simply given in this limit by $n_{\downarrow}(t) = 1-n_{\uparrow}(t)$).
As can be observed, the central level occupation is composed by contributions from the upper and lower ABS denoted by $n_\pm$ and predicts for $\Gamma\lesssim\Delta$ and this initial condition a long 
time magnetic solution. Additionally there is an 
interference term $n_{+-}$, which vanishes for the electron-hole symmetric situation ($\epsilon_0=0$). This last term is well approximate at times $t >1/\Delta$ as \cite{SNS_Souto}
\begin{equation}
n_{+-}(t) \simeq -\frac{\epsilon_0}{\pi\Delta}\left(1-\frac{\epsilon_0^2}{\epsilon_A^2}\right)\left[1+\cos(2\epsilon_At)+\sqrt{\frac{2\pi}{\Delta\cdot t}}
\cos(\epsilon_At)\sin\left(\Delta\cdot t-\frac{\pi}{4}\right)\right]
\end{equation}
\end{widetext}

For a different initial condition, $(n_\uparrow(0),n_\downarrow(0))=(0,0)$ or  $(n_\uparrow(0),n_\downarrow(0))=(1,1)$ there is a contribution from the second term in Eq. \ref{keldysh_SPA}. For an 
initially empty level the solution is non-magnetic ($n_\uparrow(t)=n_\downarrow(t)$), and we have $n_{\sigma}(t) =  \sum_{\pm} n_{\pm}(t) + n_{+-}+n_{\sigma,00}$ where
\begin{equation}
 n_{\sigma,00}(t)=2\left|p_{12}\right|^2\left[1-\cos(2\epsilon_A t)\right]\,,
\end{equation}
which describes undamped oscillations (see Fig. \ref{charge_evolution}). Similar expressions can be derived for the other initial fully occupied configuration.

\section{Interpretation in terms of rate equations}
\label{sec::rate_eq}

A simple interpretation of the numerical results for the asymptotic probabilities
$P_{\pm}(t)$ and $P_{odd}(t)=P_{odd,\uparrow}(t)+P_{odd,\downarrow}(t)$ can be obtained assuming that they are governed
by simple rate equations with time dependent rates. More precisely, these equations
are

\begin{eqnarray}
\frac{dP_-}{dt} &=& -2\Gamma_{odd}(t) P_{-}  + \Gamma_{-}(t) P_{odd} \nonumber\\
\frac{dP_{odd}}{dt} &=& 2\Gamma_{odd}(t) P_{-}  - \left[\Gamma_{-}(t)+\Gamma_{+}(t)\right] P_{odd} +2\Gamma_{odd}(t) P_{+} \nonumber\\
\frac{dP_+}{dt} &=& \Gamma_{+}(t) P_{odd}  -2\Gamma_{odd}(t) P_{+}\;,
\end{eqnarray}
where $\Gamma_{odd}(t)$ and $\Gamma_{\pm}(t)$ are the time-dependent rates for transitions 
between states. Although these quantities are not well defined, an estimate
based on perturbation theory would suggest that they should be inversely 
proportional to the energy distance from the lower gap edge to the corresponding
state. 
For $\Gamma/\Delta\gg1$ the rates can be considered as time independent 
and approximated as  $\Gamma_{odd}\approx\Gamma^2/\Delta$ and $\Gamma_{\pm}\approx\Gamma^2/(\Delta\pm\epsilon_A)$. Using these estimates one obtains
the results indicated by the dashed lines in the left panel of Fig. \ref{smoothSwitch} of the main text.

\section{Bidirectional Poisson distribution}
\label{sec::B-d}
\begin{figure}
\includegraphics[width=.7\linewidth]{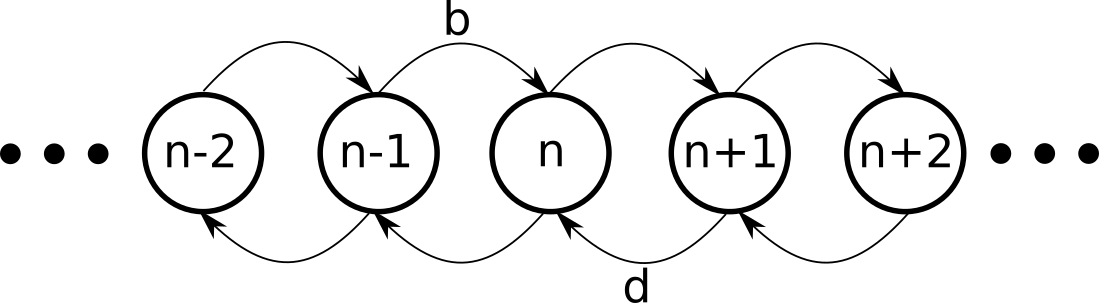}
\caption{Scheme of the birth-death process. The birth process (which has a probability $b$), allows to go from $n$ to $n+1$ charges transferred through the junction. In the short time limit, we consider
$b=d$, due to the absence of bias voltage.}
\label{birth-death_scheme}
\end{figure}

In this section we show the calculation details for the bidirectional Poisson distribution, also known in the literature as birth-death processes \cite{Gardiner_book}. 
We assume that, starting from an initial population, $n$, there is a birth rate (d)
and a death rate (d), which connects the subspaces with different population (see Fig. \ref{birth-death_scheme}). The distribution supposes that birth and death processes occur with a fixed probability,
independently from the history of the system. In our case, the population will describe the number of electrons transfer through the junction, being the initial population $n=0$. Then, a positive 
(negative) population is interpreted as a net charge flowing from the left (right) electrode to the right (left) one. The results for the short time dynamics can be found in the lower panel of Fig. 
\ref{current_sym}, where equal birth and death probability rates are considered ($b=d$).\newline

The calculation of the population is done in the following way. We start with an initial population $P(0)=(\ldots,0,1,0,\ldots)^T$. The recursive
expression for computing the probability distribution can be written as $P(t+dt)=M\cdot P(t)$, being
 \[M=\left( \begin{array}{cccc}
\ddots & \ddots &  &\\
\ddots & a & d\,dt & \\
 & d\,dt & a & \ddots\\
&  & \ddots & \ddots \end{array} \right),\]
and $a=1-(b+d)dt$ is the probability of staying at the same subspace after $dt$. The discontinuous dots mean that we consider enough probabilities to make the border effects negligible. In our problem,
the short time state of the system is fully characterize by the parameter $n_\rightarrow(t)=b\,dt=d\,dt$, which is the mean charge transfer in one of the directions of the junction (without considering 
charges flowing in the opposite direction). For fitting the Fig. \ref{short_time_FCS} at short times, we have taken $n_\rightarrow(t)\sim1$ for the discontinuous line
and $n_\rightarrow(t)\sim3$ for the dotted one. Then, the mean number of electrons crossing the junction needed necessary for the ABS to be formed is surprisingly small 
(they are of the order of $3$ electrons crossing the junction in both directions of the junction).\newline

\end{document}